\documentclass[lettersize,journal]{IEEEtran}
\usepackage{amsmath,amsfonts}
\usepackage{algorithmic}
\usepackage{algorithm}
\usepackage{array}
\usepackage[caption=false,font=normalsize,labelfont=sf,textfont=sf]{subfig}
\usepackage{textcomp}
\usepackage{stfloats}
\usepackage{url}
\usepackage{pdfpages}
\usepackage{verbatim}
\usepackage{graphicx}
\usepackage{balance} 
\usepackage{cite}
\usepackage{tabulary}
\usepackage{multirow}
\usepackage{textcomp}
\usepackage{tabularx}
\usepackage[T1]{fontenc}
\usepackage[export]{adjustbox}
\newcolumntype{M}[1]{>{\centering\arraybackslash}m{#1}}
\newcolumntype{R}[1]{>{\raggedleft\arraybackslash}m{#1}}
\newcolumntype{L}[1]{>{\raggedright\arraybackslash}m{#1}}
\usepackage{xcolor}

\usepackage{pifont}% http://ctan.org/pkg/pifont
\newcommand{\cmark}{\ding{51}}%
\usepackage{tcolorbox}
\usepackage{enumitem}
\newtcolorbox{suggestion}{breakable,colback=pink!30}
\begin{document}

\sloppy

\title{THERADIA WoZ: An Ecological Corpus for Appraisal-based Affect Research in Healthcare}
%Relying on Appraisal Theories to Build an Ecological Corpus for Multimodal Detection of Affective States in Healthcare

\author{Hippolyte Fournier, Sina Alisamir, Safaa Azzakhnini, Isabella Zsoldos, Eléonore Trân, Gérard Bailly, \\Frédéric Elisei, Béatrice Bouchot, Brice Varini, Patrick Constant, Joan Fruitet, Franck Tarpin-Bernard,\\Solange Rossato, François Portet, Olivier Koenig, Hanna Chainay, Fabien Ringeval%~\IEEEmembership{Staff,~IEEE,}

\thanks{
\textit{
H. Fournier, S. Alisamir, S. Azzakhnini, F. Portet, S. Rossato, and F. Ringeval are with the Univ. Grenoble Alpes, Inria, CNRS, Grenoble INP, LIG. E-mail:\{hippolyte.fournier, sina.alisamir, safae.azzakhnini, francois.portet, solange.rossato, fabien.ringeval\}@univ-grenoble-alpes.fr.\\
H. Chainay, O. Koenig, I. Zsoldos, and E. Trân are with the Univ. Lyon 2, EMC. E-mail:\{hanna.chainay, olivier.koenig, isabella.zsoldos, eleonore.tran\}@univ-lyon2.fr.\\
G. Bailly and F. Elisei are with the GIPSA-lab, Univ. Grenoble Alpes. E-mail:\{gerard.bailly,frederic.elisei\}@gipsa-lab.grenoble-inp.fr.\\
B. Bouchot and B. Varini are with ATOS company. E-mail:\{beatrice.bouchot,brice.varini\}@atos.net.\\
P. Constant is with Pertimm company. E-mail:\{patrick.constant\}@pertimm.com.\\
J. Fruitet and F. Tarpin-Bernard are with Humans Matter company. E-mail:\{j.fruitet, f.tarpin\}@humansmatter.co.\\}

This work has been submitted to the IEEE for possible publication. Copyright may be transferred without notice, after which this version may no longer be accessible.

}
}

        % <-this % stops a space

% <-this % stops a space

% \thanks{\textit{H. Fournier, S. Alisamir, S. Azzakhnini, F. Portet, and F. Ringeval are with the Université de Grenoble. E-mail:\{hippolyte.fournier, sina.alisamir, safae.azzakhnini, francois.portet, fabien.ringeval\}@univ-grenoble-alpes.fr.\\
% H. Chainay, O. Koenig, I. Zsoldos, and E. Trân are with the Université Lyon 2. E-mail:\{hanna.chainay, olivier.koenig, isabella.zsoldos, eleonore.tran\}@univ-lyon2.fr} \\
% Gerard 
% }

% \markboth{IEEE TRANSACTIONS ON AFFECTIVE COMPUTING,~Vol.~14, No.~8, August~2021}%

\maketitle

\begin{abstract}

We present THERADIA WoZ, an ecological corpus designed for audiovisual research on affect in healthcare. Two groups of senior individuals, consisting of 52 healthy participants and 9 individuals with Mild Cognitive Impairment (MCI), performed Computerised Cognitive Training (CCT) exercises while receiving support from a virtual assistant, tele-operated by a human in the role of a Wizard-of-Oz (WoZ). The audiovisual expressions produced by the participants were fully transcribed, and partially annotated based on dimensions derived from recent models of the appraisal theories, including novelty, intrinsic pleasantness, goal conduciveness, and coping. Additionally, the annotations included 23 affective labels drawn from the literature of achievement affects. %, 10 of which were identified as most relevant in the context of assisted CCT.
We present the protocols used for the data collection, transcription, and annotation, along with a detailed analysis of the annotated dimensions and labels. Baseline methods and results for their automatic prediction are also presented. 
The corpus aims to serve as a valuable resource for researchers in affective computing, and is made available to both industry and academia.

\end{abstract}


\section{Introduction}
\label{sec: intro}
\IEEEPARstart{O}{ver} the past decade, special attention has been given to the use of Artificial Intelligence (AI) technologies to improve healthcare. One of the rationales behind this craze is that the automation of tasks that do not necessarily require human intervention saves therapists time and mental load, resulting in more efficient practices and less burden for both patients and clinicians. However, recent reviews have evidenced that, even though AI technologies' performance can rival that of humans in certain specific tasks, such as serum analysis, or assessment of cardiovascular MRI images, they still struggle to deal with social interaction skills~\cite{YU18-AI}. 

%, JALIAAWALA20-AIAUTISM

While the definition of social interaction skills remains a subject of debate~\cite{LITTLE17-SOCIALSKILL}, they can be summarised as the ability of autonomous agents to maintain equilibrium in a dynamic relationship with another agent, resulting in the development of autonomous relationships~\cite{DEJAEGHER10-SOCIALSKILL}. In the context of interactions between AI technology and patients in healthcare, it corresponds to the capabilities of systems to detect and respond appropriately to signals from patients that may disrupt the balance of interaction; e.g., patients may express that they feel misunderstood or unsatisfied. % It is important to note that patients may convey feelings of being misunderstood through a verbal request, as well as the expression of an affective state. 
While verbally expressed requests are fairly well understood by today's AI technologies, it is still difficult for them to accurately detect patients' affective states from non-verbal signals~\cite{pepa2021automatic}.

\begin{figure}[!t]
\centering
\includegraphics[width=3.3in]{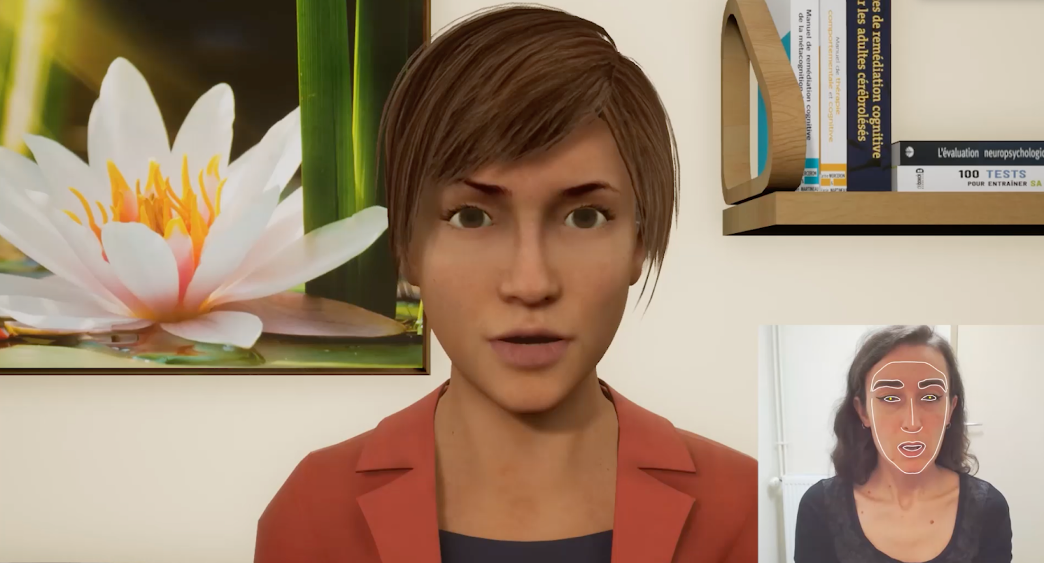}
\caption{Head movements, gaze, speech, and articulation of a remote operator were captured in real-time to drive a virtual assistant as a Wizard-of-Oz~\cite{Tarpin-Bernard21-TDT}.}
\label{fig_WoZ}
\end{figure}

Yet, the use of AI technologies for social interactions in healthcare would have far-reaching consequences. A prime illustration of their relevance lies in Computerised Cognitive Training (CCT), which aims to enhance or preserve cognitive functions in patients suffering from cognitive impairment through the repetition of exercises that target specific functions such as attention or memory~\cite{HILL17-CCT}. Although the effectiveness of CCT has been evidenced by meta-analyses~\cite{HILL17-CCT, GAVELIN22-CCT}, it seems to be conditioned by social interaction during the process~\cite{LAMPIT14-CCT}, which requires the presence of the therapist. Thus, entrusting social interaction to AI technologies would be highly beneficial in making the process autonomous. To move in this direction, it is important to use meaningful data and reliable methods to accurately recognise patients' affective state in the context of CCT. %However, existing annotated data sets of affect in the context of healthcare are rare, and do not cover the domain of CCT.%, and concern only the English language.

%LEUNG15-PARK,, HU21-CCT

In this study, we present the THERADIA WoZ corpus, an ecological corpus specifically tailored for the audiovisual detection of affective states in the healthcare domain. The corpus was built to help the development of a virtual assistant supporting CCT sessions at home, for older healthy people and patients~\cite{Tarpin-Bernard21-TDT}. To the best of our knowledge, this is the first audiovisual corpus of affect in healthcare that fully relies on models of appraisal theories, while being accessible to both industrial and academic research communities\footnote{The URL will be provided for camera-ready}.

The corpus data come from natural interactions in French involving healthy and Mild Cognitive Impairment (MCI) senior participants. The interactions revolve around CCT exercises facilitated by a virtual assistant, and operated remotely by a human acting as a Wizard-of-Oz (WoZ), cf. Figure~\ref{fig_WoZ}. 
%All participants filled in several questionnaires targeting demographic, psychological and health information. 
Expressions of the participants were then fully transcribed and partially annotated based on the dimensions of recent appraisal theories models, along with labels derived from the literature of achievement affects.%, which are contextually relevant to AI-assisted healthcare. %; this label annotation was based on continuous values, while allowing for the simultaneous presence of several labels to account for affect blends.

A comprehensive analysis of the annotated labels was carried out, enabling the identification of a core set comprising ten affective states in the context of AI technology-human interactions in healthcare. This core set of affective labels was then used along with the appraisal dimensions to evaluate the performance of inference models based on audio, visual, and textual data.

The remainder of this paper is organised as follows. We first review the appraisal theory and annotation methods of affect, along with the existing corpora in healthcare in section~\ref{sec: sota}. The protocol used for the construction of our data set is then presented with the material and the participants in section~\ref{sec: corp_const}. The segmentation and the annotation of the recordings are explained in section~\ref{sec: data_segm_annot}. A comprehensive analysis of the affective annotations 
%including affective labels and dimensions 
is then reported in section~\ref{sec: corp_a}, and results obtained in their automatic prediction are finally reported in section~\ref{sec: model}, before concluding in section~\ref{sec: conclusion}.

%Sect. 2 describes the building procedure of the virtual coach interacting environ- ment, which includes the WoZ platform, the coaching model and the preliminary studies for the agent acceptance. Sec- tion 3 describes the way in which the interaction sessions were designed and carried out and the way in which the end users were recruited. Then, in Sect. 4 the behavior of the wizard is analyzed through the language generated and the aspect of the virtual agent. Sections 5 and 6 provide a whole description of the emotional analysis of the user interactions regarding speech, language and facial expressions. S
\section{State of the art}
\label{sec: sota}

In this section, we first review the definition of affective states and their annotation methods through the lens of appraisal theories, and then provide a concise overview of corpora of affect available in healthcare.

\subsection{Affective states in the lens of the appraisal theory}
According to appraisal theories, affective phenomena are defined as multi-component processes aiming to maintain the well-being and survival of individuals in their environment \cite{SANDER13-MOD}. Roughly speaking, affective phenomena arise when the meaning of environmental stimuli are cognitively appraised as potentially threatening to an individual's equilibrium in their environment. 
%FRIJDA04-EMO, 

This appraisal process would be divided into four criteria: the relevance of the events for the individual's well-being and/or survival, their implications, the possible coping with these implications, and their normative significance, i.e., the degree of matching with individual's norms and values~\cite{SANDER05-CPM}. 

Following this appraisal process, the activation of the organism's subsystems, i.e., physiological responses, action tendencies and motor expressions, would be synchronised to help individuals carry out the behaviours that contribute to the restoration of their homeostasis.
In this perspective, an affective state serves as a marker reflecting how individuals appraise themselves between the plausible deregulation of their homeostasis caused by the environment, and the search for the behaviour leading to a return to equilibrium. 

The cognitive appraisal process is supposed to be causal in the elicitation of affective episodes~\cite{MOORS13-CAUSAL}. This position assumes a necessary and sufficient relationship would link appraisal to the synchronisation of the organism's subsystems~\cite{SIEMER07-NECESS}. In other words, a specific pattern of criteria of the appraisal process would trigger a specific synchronised activation of the organism's subsystems. Conversely, the observation of a specific synchronised activation would allow to predict the pattern of criteria of the appraisal process performed by the experimenter of the affective episode. 

Based on these premises, the Tripartite Emotion Expression and Perception model postulates that the way humans perceive the affect of others is conditioned by this causal relationship~\cite{SCHERER19-TEEP}. The idea is that if an individual $A$ is able to determine the affective state of an individual $B$, this is because $A$ inferred -- in part unconsciously -- the cognitive appraisal process performed by $B$ on the basis of the apparent motor expressions, e.g., facial expressions, that $B$ has produced. In brief, appraisal theories propose to solve the problem of affective state differentiation based on a $n$-dimensional hyperplane representing the criteria of the appraisal process. 

\subsection{Differentiable annotations of affective states}
\subsubsection{Dimensional representations}
Therefore, one way of describing individuals' affective states would be to consider them as points on a plane. Based on this approach, most of the literature have used the Circumplex model to represent them, a Cartesian plane defined by two axes, namely valence -- from unpleasant to pleasant -- and arousal -- from deactivation to activation~\cite{RUSSEL80-CIRCUM}. However, this approach drew a number of criticisms for its ability to differentiate affective states; fear and anger are both defined by high arousal and negative valence, and the concept of valence itself should be rather seen as a multidimensional representation of several dimensions~\cite{FOURNIER22-COMB}. It has therefore been stated that more than two dimensions should be taken into account to appropriately distinguish affective states~\cite{FONTAINE07-DIM}. Thus, appraisal theory represents a sound framework for selecting relevant dimensions in the context of affective state annotation.
%RUSSEL09-CORE,

\subsubsection{Categorical representations}
Labels are another common way of approximating individual's affective states. One of the crucial points of such an approach is to select affective labels likely to be relevant in the context of interest. While basic emotion labels, i.e., fear, disgust, happiness, sadness, and anger are commonly used to model affective states~\cite{TRACY11-BASIC}, they are not necessarily the most appropriate in the context of healthcare and AI-patients interactions, as they refer to affect from an evolutionary perspective.

\subsubsection{Relation between categorical and dimensional representations}
Appraisal theories suggest that the use of labels represents a way for humans to characterise the conscious part of the activation states of the organism's subsystems, i.e., of the cognitive appraisal process, the physiological responses, the action tendencies, and the motor expressions~\cite{GRANDJEAN08-CONSCIOUS}, assuming that each affective label would represent a specific pattern of criteria of the appraisal process~\cite{SCHERER19-TEEP, SCHERER18-EMOTINFER}.
%MOORS14-FLAVOR, SCHERER13-APPLAB

\subsection{Affect corpora for healthcare}
Automatic affect analysis can be applied to various aspects of healthcare~\cite{LOW20-AFFhealthcare}: to support clinicians for the diagnosis of affective disorders, to facilitate the monitoring of patients' emotions during the therapy~\cite{Tarpin-Bernard21-TDT}, which is the focus of this study, or between therapy sessions.
%~\cite{Ringeval19-A2W}, ~\cite{ADIKARI2022318}

Whereas numerous audiovisual corpora are available in the literature for affect recognition in human interactions~\cite{KOSSAIFI19-SEWA}, few of them were acquired in the context of healthcare. %, and even fewer with regard to supporting clinicians during therapy. 
There are several reasons for this gap in the literature. One of them concerns the sensitive nature of the data that are protected by law and cannot be easily released. Moreover, populations related to healthcare are often fragile, and setting up experimental protocols to contrast affect corpora can be challenging because of their cognitive fatigue.

Nevertheless, there have been initiatives in the construction of affective corpora supporting clinicians during therapy. For instance, speech interactions between seniors and a virtual assistant playing the role of a health professional coach, were collected for the purpose of the EMPATHIC project~\cite{PALMEIRO23-EMPHATICSPEECH}, which aims to %develop a virtual assistant to 
help the elderly maintain their independence as they age. Authors collected $\sim$13k speech segments ($\sim$13h in total) from 136 participants over 64 years and from different countries (Spain, France, and Norway), that were further annotated by three annotators according to five labels: calm, happy, puzzled, tense, and sad, and three dimensions: arousal, valence, and dominance. %Authors reported a strong imbalance in the frequency of the reported emotion labels, with ``sad'' and ``tense'' being present only on 17 and 14 audio segments, respectively, while the emotion ``calm'' covers more than 93\% of the audio segments. ~\cite{OLASO21-EMPATHIC}

Another corpus of affect has recently been collected for children with developmental disorders using a mobile therapeutic game~\cite{WASHINGTON21-GUESS}. The corpus contains $\sim$2k video sequences of 456 children from all over the world and was annotated by 11 individuals according to seven labels: happy, surprised, sad, fearful, angry, disgust, and neutral.

\section{Data collection}
\label{sec: corp_const}
The whole data collection protocol of this study has been documented in detail, reviewed, and approved by the Ethics Committee for Research in Grenoble Alpes with the reference: CERGA-Avis-2021-1. As the experiment lasted around three hours on average, and participants were asked to consent for the commercial exploitation of their data under the protection of the European General Data Protection Regulation laws, a 20€ gift card was awarded. Participants could ask for reimbursement of travel expenses, i.e., including public transport fares or petrol costs, for both the journey to and from the experiment site. The protocol used to collect data for the THERADIA WoZ corpus is detailed below.%, which includes the methods used for the collection, transcription, and annotation of the data. 

%\subsection{Ethical review}
%, especially the protocol used for the recruitment of the participants, the questionnaires and the consent forms given to them for the processing and sharing of their audiovisual data, 

\subsection{Participants}
A total of 52 healthy senior participants (40 females; mean age = 67.9, SD age = 5.1) were recruited based on advertisements published in regional newspapers such as ``Le Progrès'' and ``Le Dauphiné Libéré''. Additionnaly, 9 participants diagnosed with MCI (two females; mean age = 75.8, SD age = 2.1) were recruited via clinicians collaborating on the project. Inclusion criteria included fluency in French, either normal or corrected-to-normal vision and hearing, and informed consent prior to recordings. Regarding the level of education, 40\% of healthy senior participants had a master's degree or higher, 33\% a bachelor's degree and 27\% a baccalaureate. As for MCI participants, 22\% had obtained a master's degree or higher, 22\% a bachelor's degree, and 56\% a National Vocational Qualification (French CAP), or lower. 
% Moreover, the MCI participants represent only 8\,\% of the annotated data, despite the fact that the THERADIA-WoZ corpus is focused on helping people with cognitive disorders. However, this turns out to not pose a problem, as a study of the bias and fairness of the trained model was also carried out, which showed that training machine learning models with mostly healthy senior participants is indeed generalisable to the MCI population.

\begin{figure}[!t]
\centering
\includegraphics[width=2.8in]{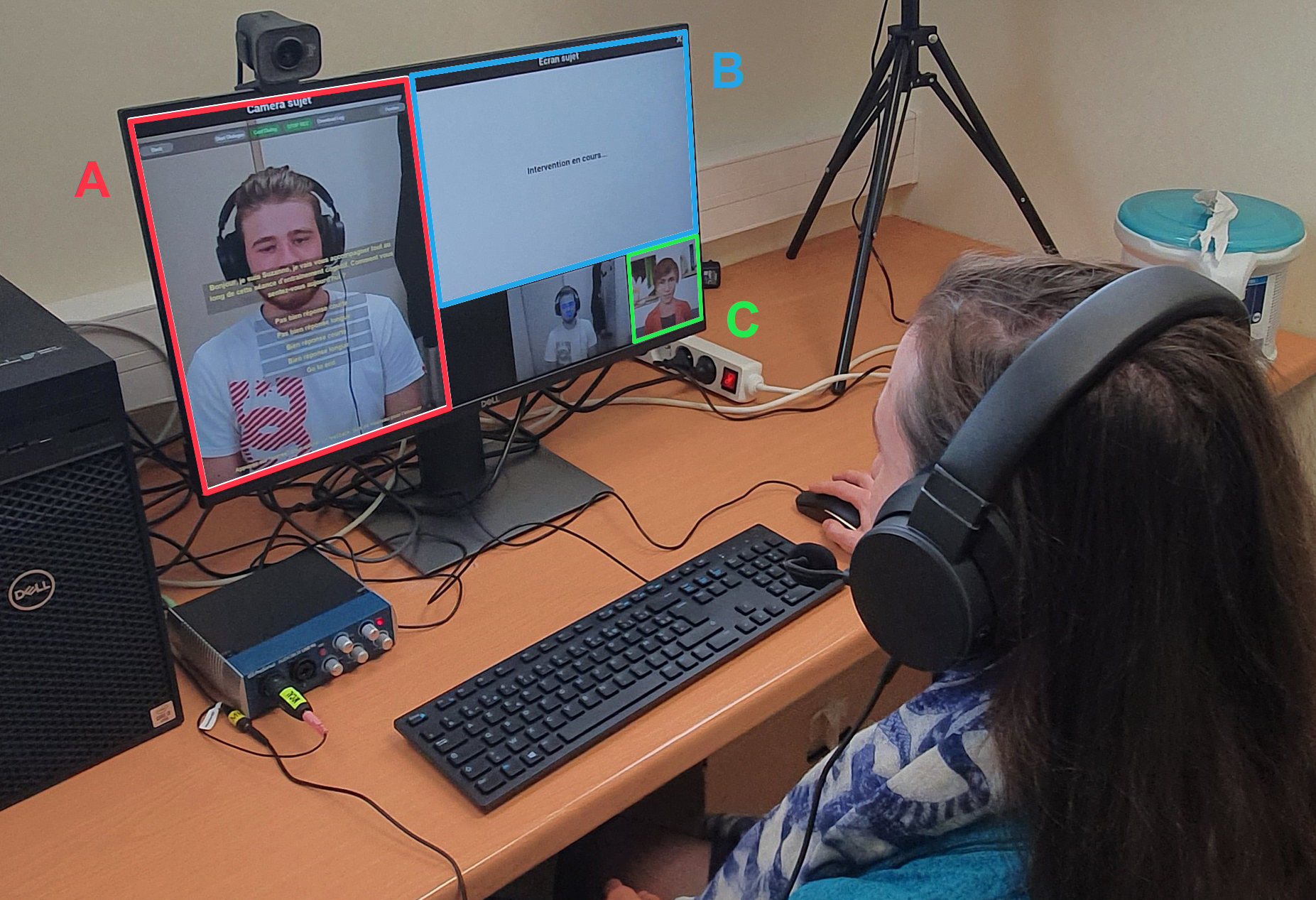}
\caption{Office of the experimenter operating the virtual assistant. The dialog was fully scripted and overlaid onto the participant's video (A). %; options were also offered to specify their feedback, e.g., ``Ok'' vs. ``Sceptical''. 
The participant's screen (B) was only visible during the exercise completion, and the virtual assistant was visible to the participant only outside the exercise completion. The virtual assistant's rendering was consistently displayed (C) to monitor any tracking issue.}
\label{Dispositif_WoZ}
\end{figure}

\subsection{CCT supported by a Virtual Assistant}
All participants carried out a CCT session, which was based on eight exercises selected from the HappyNeuronPro CCT platform. %\footnote{\url{https://www.happyneuron.fr}}
The exercises involved different cognitive functions such as memory, language, attention, and planning; details on these exercises are provided in section I.A of the supplementary material. 

The CCT was supported by a virtual assistant, developed by the Dynamixyz company%\footnote{Dynamixyz has been acquired by Take-Two Interactive in 2021.}
, and controlled by one co-author of this study acting as a WoZ, cf. Figure~\ref{Dispositif_WoZ}. More specifically, head movements, gaze, speech, and articulation of the experimenter were captured in real-time to drive the 3D virtual assistant whose rendering was cast to the participant's screen. The speech was based on a dialogue tree designed to cover a wide range of interactions, envisioned prior to the study. If none of the planned branches made sense within a given dialogue, free intervention was allowed. Eventually, the dialogue tree was updated if the unplanned interaction was deemed relevant. Technical problems that could not be solved by the operator of the virtual assistant were addressed by a second experimenter, who was sitting outside the room for the healthy participants, and inside the room for the MCI participants. 
% working directly in the participant's room; the experimenter

\subsection{Material}
Questionnaires were used to interrogate various psychological, affective and cognitive dimensions of the participants, as well as to evaluate the virtual assistant and the CCT~\cite{ZSOLDOS-SUB-WOZ}. Experiments were carried out in two quiet rooms of the EMC laboratory of Université Lyon 2, one for the participant, and one for the experimenter.% as illustrated in the Figure~\ref{fig_design}.

As the original lighting conditions did not allow a robust tracking of the operator of the virtual assistant, we equipped each room with an additional continuous light (Amzdeal Softbox) that provided a color temperature closed to daylight (5500K). Each room was equipped with an high-end computer, comprising a 24 inches monitor, a webcam (Logitech streamcam), a headset with a built-in unidirectional microphone (Sound blasterX H6), that was plugged to an external sound card (Presonus Audiobox), for both audio recording and streaming to the CCT application. 

As pre-tests revealed random freezes in the video recordings that were not apparent during the CCT session, we used an iPhone X to provide a stable video recording on the participant's side, along with an additional audio recording captured by a far-field omnidirectional microphone. Audio data were sampled at 44.1 kHz with 16 bits, whereas video data were sampled at a constant frame rate of 30 fps with a resolution of 1920x1080 and the YUV420p color format.

\subsection{Procedure}
%Participant and the WoZ were housed in the two separate experimental rooms. 
Concerning the MCI participants, the session began with the completion of the consent form and the above-mentioned questionnaires. %, with the exception of the debriefing questionnaire evaluating the protocol. 
Then, after a few instructions, the experimenter left the room so that the CCT session could start. It comprised the realisation of four exercises, each performed twice. The difficulty level of the second exercise was adjusted based on the performance of the first. It was increased in the case of success and decreased in the case of difficulty. Throughout the session, participants were guided and engaged with the virtual assistant under the impression it was operated using AI technology, whereas it was actually operated by a human as a WoZ. The story line of the session started with a welcome dialogue designed to acquaint the participant with the virtual assistant, and to collect some information such as the participant's mood and motivation. This was followed by the exercise phase divided into five parts: the introductory dialogue detailing each exercise, the exercise completion, the feedback dialogue commenting on participant's performance, the repetition of the exercise, as well as the feedback dialogue, and the closing dialogue, which ended the session. Once the CCT session was terminated, the whole set-up was revealed.

Concerning the healthy senior participants, the procedure was the same, except for the following: First, a session consisted of eight exercises instead of four. Second, healthy senior participants were divided into five groups. Twenty of them were assigned to a group without induction. The special feature of this group was that it carried out two CCT sessions, one week apart. This resulted in the set-up being revealed at the end of the second session. The remaining 32 healthy senior participants were divided into four induction groups, designed to increase the chances of observing affectively-charged expressions. To this end, induction was used in one out of two of the eight exercises based on the combination of three parameters: (i) presenting the exercise as easy or hard beforehand; (ii) actually setting the difficulty as easy or hard; and (iii) giving feedback from the WoZ as critical or congratulatory. Details on the groups and the inductions are reported in section I.B of the supplementary material.

\begin{table}[t]
\setlength{\tabcolsep}{9 pt}
\renewcommand{\arraystretch}{1.6}
\scriptsize
\caption{Overall statistics of the data collected in the THERADIA-Woz}%The overall duration is given in hours:minutes:seconds format.
\label{tab:corpus-stats}
\begin{tabular}{llrrr}
\hline
Data & \begin{tabular}[c]{@{}l@{}}Participants\end{tabular} & \multicolumn{1}{c}{\begin{tabular}[c]{@{}c@{}}\# Seq.\end{tabular}} & \multicolumn{1}{c}{\begin{tabular}[c]{@{}c@{}}Seq. duration\\ Mean (std)\end{tabular}} & \multicolumn{1}{c}{\begin{tabular}[c]{@{}c@{}}Overall \\ duration\end{tabular}} \\ \hline
\multirow{3}{*}{Transcribed} & \textbf{Senior}& 14,788& 9.00\,s (7.82\,s) & 36h59m16s \\ %\cline{2-5} 
& \textbf{MCI}& 1,105& 8.24\,s (5.03\,s) & 2h31m44s \\ %\cline{2-5} 
& \textbf{Total}& 15,893& 8.95\,s (7.66\,s) & 39h31m00s \\ \hline
\multirow{3}{*}{Annotated} & \textbf{Senior}& 2,513& 8.20\,s (5.27\,s) & 5h43m27s \\ %\cline{2-5} 
& \textbf{MCI}& 222& 10.09\,s (6.49\,s) & 37m20s \\ %\cline{2-5} 
& \textbf{Total}& 2,735& 8.35\,s (5.40\,s) & 6h20m47s \\ 
\hline
\end{tabular}
\end{table}

%\tiny, \scriptsize, \footnotesize, \small, \normalsize, \large, \Large, \LARGE, \huge, \Huge

% \hline
% \textbf{All-Senior} & 71 sessions & 79h54m24s & 1h07m33s (11m35s) \\
% \textbf{All-MCI} & 9 sessions & 5h59m09s & 39m54s (11m45s) \\
% \textbf{All-Total} & 162 sessions & 133h09m56s & 1h00m32s (12m35s) \\

%\begin{table*}[t]
%\small
%  \caption{Overall statistics of the data collected in the THERADIA-Woz. The duration is given in hours:minutes:seconds format. \textbf{Transpose it, and put other results, also mean +- std}}
%  \label{tab:corpus-stats}
%  \centering
  
%  \begin{tabulary}{1.0\textwidth}{M{0.15\textwidth}|M{0.1\textwidth}|M{0.1\textwidth}|M{0.15\textwidth}}
%Groups & Number of sequences & Overall duration & Mean sequence duration (std) \\ 

%\hline
%\textbf{Transcribed-Senior} & 14788 & 36h59m16s & 9.00 s (7.82 s) \\
%\textbf{Transcribed-MCI} & 1105 & 2h31m44s & 8.24 s	(5.03 s) \\
%\textbf{Transcribed-Total} & 15893 & 39h31m00s & 8.95 s	(7.66 s) \\

%\hline
%\textbf{Annotated-Senior} & 2504 & 5h42m33s & 8.21 s (5.27 s) \\
%\textbf{Annotated-MCI} & 222 & 37m20s & 10.09 s (6.49 s) \\
%\textbf{Annotated-Total} & 2726 & 6h19m54s & 8.36 s (5.40 s) \\
        
%\end{tabulary}
%\end{table*}

\section{Data segmentation and annotation}
\label{sec: data_segm_annot}
%In total, 
We collected about 80 hours of data from the 52 healthy older partipants, and about 6 hours of data from the 9 MCI participants. We describe in this section the methods used for the segmentation, transcription, and annotation of this data.

\subsection{Segmentation, transcription and selection of sequences}
%The main objective was to provide a meaningful segmentation of the recordings, based on both verbal and non-verbal information. 
The segmentation of the recordings was based on the pragmatic completeness in breath groups, i.e., groups of words that are delimited by pauses taken for breathing, ensuring they formed a single proposition if they conveyed the same semantic information, or separate propositions if they related to different types of semantic information~\cite{ford1996interactional}. Non-verbal expressions occurring within or proximate to a proposition were marked by a diacritic, and those occurring at distance were specifically segmented. %Those non-verbal expressions were conveyed by either speech sounds that could not be transcribed, or facial and/or gestural expressions. 
Temporal markers for proposition start and stop were both positioned at a certain distance to allow a sufficient observation space for the annotation of affect.%, especially for the proposition start. 

The transcription relied mostly on the ESLO project principles~\cite{eshkol2011grand}, respecting spelling and spoken structures; details regarding our conventions are reported in section II.A of the supplementary material. Recordings were transcribed with the ELAN software~\cite{brugman2004annotating} by five Master students in Language Engineering at the Université Grenoble Alpes with a two-steps process: an initial stage of segmentation and transcription, followed by a subsequent verification stage conducted by another student. The segmented sequences were selected to retain only those containing expressions of affect, and a fraction was randomly removed to reduce the overall duration. Obtained sequences were then all reviewed and some were further removed to reduce redundancy. Statistics of the sequences are given in table \ref{tab:corpus-stats}.

\subsection{Selection of affective labels and dimensions}
The context of CCT aligns closely with the literature on achievement affects~\cite{PEKRUN06-ACHIEV,PEKRUN17-ACHIEV,SILVIA10-CURIOS}%, which focuses on the most relevant affects in the context of the school environment. % in which students are involved in completing exercises with the help of a teacher. 
%PEKRUN12-ACHIEV,PEKRUN10-ACHIEV, ,
%According to this theoretical framework,
, which considers the following types of affect: 
%where the most experimented affects %in this context 
%can be classified as follows: 
\begin{enumerate}[label=(\roman*)]
\item activity affects generated by the task achievement itself;
\item outcome affects generated by the success or failure results, whether retrospective or prospective;
\item epistemic affects generated by a form of metacognition on the perception of the achievement;
\item social affects generated by another individual.
\end{enumerate}

%From our point of view, 

These affects are particularly relevant in the context of AI-assisted healthcare, as they comprehensively cover those that can be experienced during CCT, i.e., those generated by the CCT itself (activity affects), those generated by past or future CCT failure or success (outcome affects), those generated by the perception of the cognitive abilities involved in CCT (epistemic affects), and those generated by AI-patients interactions (social affects). Our categorical representations of affect relied on the following 23 labels: %related to achievement affects: 
angry, annoyed, anxious, ashamed, confident, contemptuous, curious, desperate, disappointed, embarrassed, excited, frustrated, interested, guilty, happy, hopeful, impatient, proud, relaxed, sad, satisfied, surprised, and upset.

The dimensions consisted of the four most important appraisal dimensions in the context of affect recognition~\cite{SCHERER19-TEEP,SCHERER21-INVEST, SCHERER13-APPLAB, SCHERER18-EMOTINFER}, namely: novelty, intrinsic pleasantness, goal conduciveness, and coping, plus arousal. In the context of appraisal theory, arousal is a marker of the depth of the appraisal process, which is relevant to qualify the intensity of the affective state. 

\subsection{Annotation tool and guidelines}

%which is based on the Angular framework for the front-end, and the python library for the back-end,
We took benefit from previous collaborations with the company Viadialog\footnote{\url{https://viadialog.com/}}
engaged in the development of a web-based annotation tool originally tailored for audio data: ANNOT. This tool has been adapted to the needs of our project and is available online\footnote{The URL will be provided for camera-ready}. The platform allowed annotators -- here 6 French students (3 females) from different backgrounds (engineering, business, and biology) and places (Lyon and Toulouse) -- to view and annotate the selected audiovisual sequences at home, using their own computer.

\begin{figure}[!t]
\centering
\includegraphics[width=3.4in]{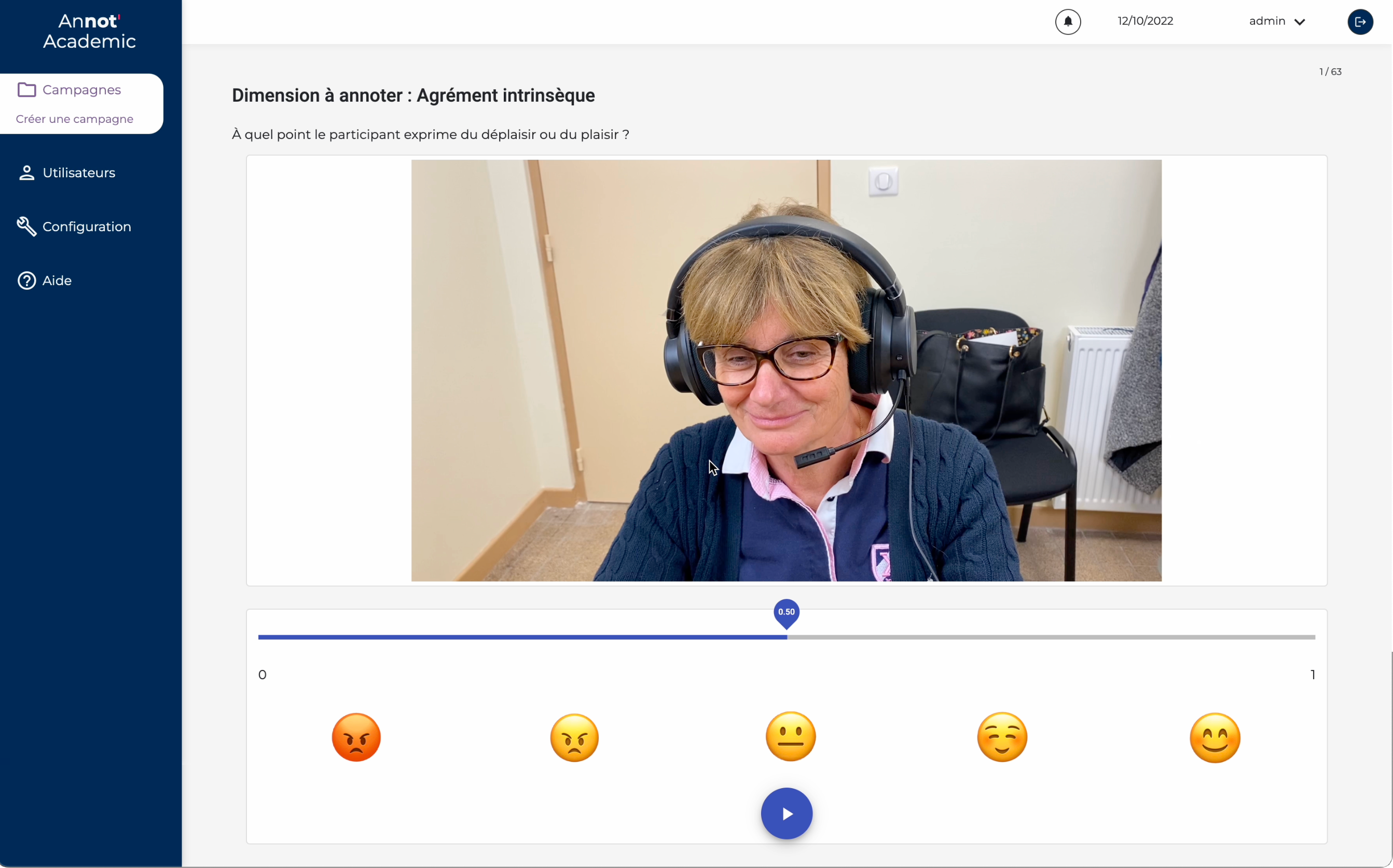}
\caption{Example of the ANNOT web-base platform for time-continuous annotation of the intrinsic pleasantness dimension.}
\label{Annot}
\end{figure}

The annotation consisted of three steps: a time-continuous annotation, cf. Figure~\ref{Annot}, followed by a summary value, for each of the five dimensions, and a summary value of intensity for one or multiple labels. Thus, each sequence was viewed five times, once for each dimension. At the end of the time-continuous annotation, a graph showing the dynamics of the dimension over time was presented to the annotators, who could then validate or perform the annotation again. After the validation, annotators were asked to provide a summary value for the whole sequence. Finally, once all dimensions were annotated, annotators were asked to indicate if one or more labels could describe the affect they had recognised in the sequence and provide an intensity value for them. Annotators could annotate as many labels as desired, or none at all if the sequence lacked any particular affect. Further details on the annotations steps are reported in section II.C of the supplementary material.%; values ranged from 0 to 1 (with a step of 0.01), with neutrality set at 0.5 for the dimensions, and from 0 (default), absent to 100, fully present (with a step of 1), for the labels' intensity. 

%In this last annotation part, annotators could review the sequence and pause it if they wished.

%, and a version can also be find as a software-as-a-service\footnote{The URL will be provided for camera-ready}.%https://github.com/voicelab-org/labelit
%The Annot' platform (https://annot-theradia.imag.fr/) has been specifically developed for annotation as part of the THERADIA project. This platform, hosted online by Viadialog, enabled annotators to view and annotate extracts as appropriate to project needs at home on their own computer. More precisely, it consisted of three types of annotations: a continuous time annotation at the same time of the extract viewing for five dimensions; a summary value annotation for each dimension; and a summary value annotation of one or multiple labels. 

%Each annotator was provided with a headset. 

Before starting, annotators participated in a training session with the project investigators, including an expert in appraisal theories of emotion. The purpose of this session was to familiarise the annotators with the concepts of appraisal theories, i.e., the meaning of dimensions and affective labels, and to explain in detail the guidelines for the annotation process. Annotators were instructed to put themselves in the participant’s shoes to assess the emotional expression conveyed in the selected sequences, considering factors such as the chosen words, their oral delivery, and the expressive dynamics of the face and body. It was emphasised that there were no right or wrong answers, as the task was inherently subjective. Additionally, they were instructed to perform the task in full-screen mode, in a quiet environment, with sessions lasting no longer than one hour, followed by ten-minute breaks. The annotation process was continuously monitored through an online tracking table, updated daily by the annotators, with additional communication via email or phone as needed. After the training session, each annotator received a document 
% Fabien: please be careful once again, we use British English, not US, so do not use z instead of s (summarizing -> summarising)
summarising all these points, including definitions of dimensions and labels. Each dimension was defined as follows:

%Paragraphe avant modification pour reviewws :
%A document was provided for each annotator as guidelines. It was explained that the annotation task had to be performed in full screen mode in a silent place. The document advised against annotating too many sequences consecutively and suggested a ten-minute break after an hour of work. We also stressed that there were no right or wrong answers in the annotation, as the information to be judged was subjective in nature. Thus, the annotations were to be made by trying to put oneself in the participant's shoes to assess his or her feeling about the expression produced in the selected sequences. These expressions could include the chosen words, the way they were communicated orally, and the expressive dynamics of the face and body. Finally, the document provided definitions about dimensions and labels. Each dimension was defined as follows:
\begin{itemize}
\item \textit{Novelty}: ``To what extent does the participant feel that what is happening is predictable or unexpected?''; ranging from \textit{predictable} to \textit{unexpected}.
\item \textit{Intrinsic pleasantness}: ``At what point does the participant express pleasure or displeasure?''; ranging from \textit{unpleasantness} to \textit{pleasantness}.
\item \textit{Goal conduciveness}: ``How much does what is happening seems to correspond to the participant's wishes?''; ranging from \textit{not at all} to \textit{absolutely}.
\item \textit{Coping}: ``How well does the participant seem able to cope with what is happening?''; ranging from \textit{not at all} to \textit{absolutely}.
\item \textit{Arousal}: ``How asleep or awake does the participant seem?''; ranging from \textit{asleep} to \textit{awake}. 
\end{itemize}

%After the annotators had been given several days to familiarize themselves with the document, they met collaborators of the project explaining them again the protocol and the definition of dimensions and labels, and answered their questions.

\section{Corpus Analysis}
\label{sec: corp_a}
This section provides a comprehensive statistical analysis of the annotation data. We first assessed the annotation frequency of the affective labels as well as the inter-annotator agreement on their intensity. We then evaluated the cognitive delay involved in the annotation of each dimension, as well as the inter-annotator agreement. Relationships between dimensions and affective labels were also estimated, and a label core set was finally selected, based on compliance with consistency criteria relating to each of the analyses mentioned above.

%A crucial point for experiments involving categorical annotations was to propose the most likely relevant labels for the context. To address this challenge, we voluntarily proposed a large number of affective labels to identify the most relevant in the context of WoZ-assisted CCT. 

\begin{figure*}[t!]
%\centering
\includegraphics[width=.95\textwidth]{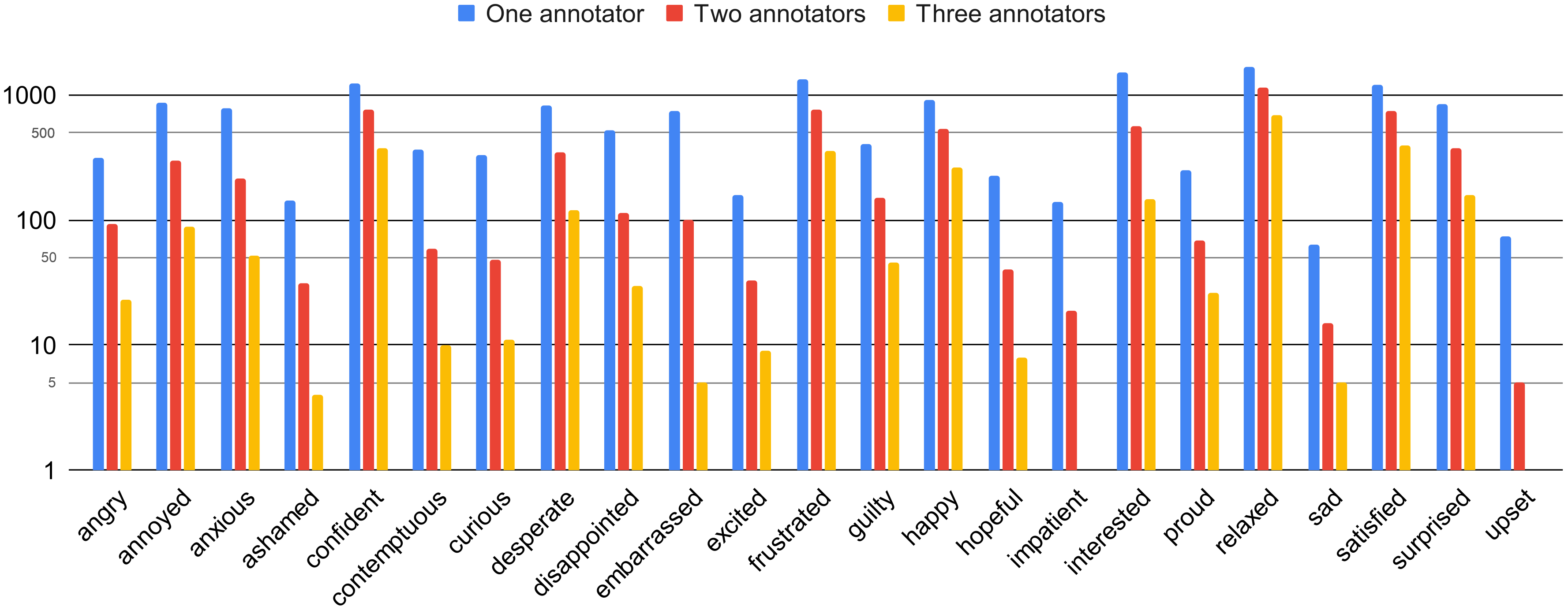}
\caption{Frequency (logarithmic scale) of the annotated labels according to the agreement of at least one, two or three annotators among six.}%Note that the vertical axis is on a logarithmic scale for visual purposes.
\label{fig:freq_labels}
\end{figure*}

\subsection{Categorical annotations}
\label{sec:label_analysis}

The frequency at which the affective labels were selected by at least one, two, or three annotators is reported in Figure~\ref{fig:freq_labels}. Based on the sample of annotation frequency per label for at least one annotator, Cohen’s $d$ was calculated to identify the theoretical value at which the effect size of its deviation from the sample mean can be considered at least "small", i.e., Cohen's $d$ > 0.2 ~\cite{cohen2016power}; this theoretical value was equal to 747 sequences. The annotation frequency per label for at least one annotator is reported in Table \ref{tab:coreset-overview}, and shown in bold when above this theoretical value.

%The following ten affective labels were annotated more frequently than this value: “Confident”, “Happy”, “Interested”, “Relaxed”, “Satisfied”, Annoyed”, “Anxious”, “Desperate”, “Frustrated”, “Surprised”. 

%The most frequently observed and thus relevant affects in the context of AI-assisted CCT session contained not only positive labels such as ``relaxed'', ``interested'', ``confident'', and ``satisfied'', but also negative labels such as ``frustrated'', ``annoyed'', ``desperate'', and ``anxious''. The least frequently reported affects are mostly associated with negative labels: ``sad'', ``upset'', ``impatient'', and ``ashamed''. 

%Whereas some labels were rarely observed in the data, such as ashamed, impatient, upset, and sad

We evaluated the inter-annotator agreement on the presence of each affect label, i.e., when the intensity of a label was superior to zero, for each pair of annotators, and averaged over all pairs, by computing the Unweighted Average Recall (UAR) as follows:
\begin{equation}
    UAR = \frac{1}{k} \sum_{i=1}^k \frac{N_c^i}{N^i}
    \label{UAR}
\end{equation}where $k$ is the total number of classes, which corresponds here to the presence or absence of a label. $N^i$ is the total number of annotated sequences for a given affect label, and $N_c^i$ is the number of commonly identified classes, whether present or absent. Results are reported in Table \ref{tab:coreset-overview}.
%Note that if the number of instances in different classes is balanced, then the UAR is equal to the accuracy. %Therefore, UAR has also been mentioned in the literature as "weighted accuracy". 
According to a two-tailored $z$-test, the inter-annotator agreement would be defined as statistically above chance if it would exceed 53\%, which is the case for all the labels excepted embarrassed.%; those labels are marked in bold style in Table \ref{tab:coreset-overview}.

\begin{table}[t]
\small

  \caption{Statistics on the annotated affective labels: frequency, inter-annotator agreement on the presence, and intensity, and correlation to appraisal's dimensions. Most relevant labels in the context of AI-assisted CCT are marked in bold} %Affective label annotations coreset (bolded) selection overview. Labels are sorted based on frequency and inclusion in the core set was satisfied if at least three out of the four criteria related to frequency (top ten), agreement on the label ($>$60\,\%), agreement on the intensity of the label ($>$.1), and the overall correlation with the appraisal dimensions, were fulfilled. }%: 1) the emotion label annotation frequency is in the top ten or not, 2) the inter-annotator agreement (UAR) of the presence of the label is bigger than 60\%, 3) the inter-annotator agreement (CCC) of the intensity of the label is bigger than .1\%, and 4) there is a relationship between the dimensional annotations and the annotated label
  \label{tab:coreset-overview}
  \centering

%\begin{tabulary}{\textwidth}{c|c|c|c|c}
%\begin{tabulary}{1.0\textwidth}{m{0.07\textwidth}R{0.07\textwidth}R{0.07\textwidth}R{0.05\textwidth}M{0.08\textwidth}}
%\hline
%\multirow{2}{*}{Label} & \multirow{2}{*}{\multicolumn{1}{c}{Frequency}} & \multicolumn{2}{c|}{Agreement on} & \multicolumn{1}{c}{Correlation to} \\
%&  & \multicolumn{1}{c}{Presence} & \multicolumn{1}{c|}{Intensity} & \multicolumn{1}{c}{Appraisals} \\
\begin{tabulary}{\linewidth}{lcccc}
\hline
\rule{-3pt}{3ex}
\multirow{2}{*}{Label} & \multirow{2}{*}{Frequency} & \multicolumn{2}{c}{Agreement on} & \multirow{2}{*}{Cor. to} \\
\rule{0pt}{3ex}
 &  & Presence & Intensity & Appraisals \\
\hline 
\rule{-3pt}{3ex}
\textbf{Relaxed} & \textbf{1683} & \textbf{59.24\,\%} & \textbf{.114} & \cmark \\ %\hline
\textbf{Interested} & \textbf{1525} & \textbf{54.99\,\%} & \textbf{.194} & \cmark \\ %\hline
\textbf{Frustrated} & \textbf{1315} & \textbf{59.05\,\%} & \textbf{.211} & \cmark \\ %\hline
\textbf{Satisfied} & \textbf{1216} & \textbf{60.69\,\%} & \textbf{.097} & \cmark\\ %\hline
\textbf{Confident} & \textbf{1238} & \textbf{60.97\,\%} & \textbf{.116} & \cmark \\ %\hline
\textbf{Happy} & \textbf{902} & \textbf{62.81\,\%} & \textbf{.280} & \cmark \\ %\hline
\textbf{Surprised} & \textbf{851} & \textbf{58.44\,\%} & \textbf{.260} & \cmark \\ %\hline
\textbf{Annoyed} & \textbf{856} & \textbf{57.52\,\%} & \textbf{.130} & \cmark \\ %\hline
\textbf{Desperate} & \textbf{826} & \textbf{58.99\,\%} & \textbf{.276} & \cmark\\ %\hline
\textbf{Anxious} & \textbf{783} & \textbf{54.53\,\%} & \textbf{.283} & \cmark \\ %\hline
%\hline
Embarrassed & 745 & 51.43\,\% & -.256 & \\ %\hline
Disappointed & 520 & \textbf{56.00\,\%} & \textbf{.387} & \\ %\hline
Guilty & 400& \textbf{57.44\,\%} & \textbf{.129} & \cmark \\ %\hline
Contemptuous & 363& \textbf{53.61\,\%} & \textbf{.359} &  \\ %\hline
Angry & 311 & \textbf{56.76\,\%} & \textbf{.487} & \cmark\\ %\hline
Curious & 328 & \textbf{55.60\,\%} & \textbf{.439} & \\ %\hline
Hopeful & 223 & \textbf{55.62\,\%} & \textbf{.480} & \cmark \\ %\hline
Proud & 249 & \textbf{58.79\,\%} & .015 & \\ %\hline
Excited & 159 & \textbf{56.39\,\%} & \textbf{.150} &  \\ %\hline
Ashamed & 144 & \textbf{54.22\,\%} & -.014 & \\ %\hline
Impatient & 138  & \textbf{53.14\,\%} & -.338 & \\ %\hline
Upset & 74 & \textbf{53.48\,\%} & \textbf{1.00} &  \\ %\hline
Sad & 63 & \textbf{57.28\,\%} & \textbf{.668} &  \\ %\hline

\hline
        
\end{tabulary}
\end{table}

We further evaluated the inter-rater agreement on the intensity of the affective labels using the Pearson's Correlation Coefficient (PCC). The inter-rater agreement was defined as sufficiently high if its Cohen's $d$, i.e., its PCC value divided by the sample standard deviation, is superior to .2~\cite{cohen2016power}, which corresponds to a PCC greater than .065. This concerned most of the affective labels, which are reported in bold style. %in Table \ref{tab:coreset-overview}.%, excepted ashamed, embarrassed, impatient, and proud.

\begin{table}[t!]
\centering
\setlength{\tabcolsep}{15pt}
\renewcommand{\arraystretch}{1.6}
\footnotesize
\caption{Inter-annotator agreement on each sequence, or CCT session, computed with the Pearson's correlation coefficient for both time-continuous and summary values of appraisal dimensions; mean (standard-deviation)}
\label{tab:agreement_dim}

\begin{tabular}{lll}
\hline
                       & \multicolumn{1}{c}{Sequence}    & \multicolumn{1}{c}{Session}   \\ \hline
\multicolumn{3}{l}{Time-continuous values}                                                      \\ \hline
Arousal                & .381  (.034) & .425  (.060) \\
Coping                 & .586  (.026) & .573  (.032) \\
Goal conduciveness    & .602  (.019) & .539  (.031) \\
Intrinsic pleasantness & .588  (.088) & .547  (.038) \\
Novelty                & .307  (.088) & .294  (.128)  \\ \hline
\multicolumn{3}{l}{Summary values}                                                       \\ \hline
Arousal                & .248  (.039) & .228  (.065) \\
Coping                 & .679  (.024) & .676  (.038) \\
Goal conduciveness    & .695  (.015) & .680  (.032) \\
Intrinsic pleasantness & .662  (.020) & .619  (.036) \\
Novelty                & .401  (.113) & .403  (.150) \\ 
\hline
\end{tabular}
\end{table}

\subsection{Dimensional annotations}
\label{dim-annot}

The cognitive delay in the continuous annotation of each dimension was evaluated, and results showed that goal conduciveness (3.3\,seconds) had a significantly longer delay than the other dimensions (less than 2 seconds), suggesting that annotators had more difficulty assessing it. This evaluation is detailed in section III.A of the supplementary material.

We assessed the inter-annotator agreement on the dimensions for both time-continuous and summary annotations, using the PCC, which was computed either on each sequence, or each CCT session, cf. Table \ref{tab:agreement_dim}. An high agreement was observed on the dimensions of coping, goal conduciveness, and intrinsic pleasantness, with a more pronounced consistency observed in summary annotations compared to time-continuous annotations, which is expected since time-continuous data imply a greater variation in the possible outcomes. A lower agreement was however found for arousal and novelty, which might be explained by several reasons as detailed below. %, both for time-continuous and summary annotations This lower agreement can be explained by the fact that these two dimensions were probably more difficult to annotate in the context of our study, for reasons of their own.

The novelty dimension was assumed to be more difficult to annotate, as it probably concerned very specific cases of our study, e.g., unplanned feedback from the virtual assistant on the participant's performance, occurring in fewer sequences than the other dimensions. This assumption was corroborated by an analysis evidencing that the frequency of high-value annotations was lower for novelty compared to other dimensions; for a high-value threshold set at .85, the proportion of high-value annotations for novelty accounted for only .13\% of the time-continuous annotation data, whereas other dimensions were at least twice as frequent: .30\% for arousal, .24\% for intrinsic pleasantness, .62\% for goal conduciveness, and .36\% for coping. A similar trend was observed in summary annotations, with the occurrence of high-value annotations for novelty accounting to .26\%, whereas all other dimensions were twice as prevalent; arousal: .92\%, intrinsic pleasantness: .52\%, goal conduciveness: 1.35\%, and coping: 2.16\%.

%This assumption is corroborated with an analysis evidencing that the number of high-value annotations is lower for novelty than for the other dimensions; for a threshold defined as greater than .85, the number of high-value annotations for novelty concerned only .13\% of time-continuous annotation data while other dimensions were at least about twice more numerous; .30\% for arousal, .24\% for intrinsic pleasantness, .62\% for goal conduciveness, and .36\% for coping. The same effect was observed on summary annotations, with a number of high-value annotations for novelty equal to .26\%, and all other dimensions being at least twice more frequent; arousal: .92\%, intrinsic pleasantness: .52\%, goal conduciveness: 1.35\%, and coping: 2.16\%.

\begin{figure}[t!]
\centering
\includegraphics[width=.47 \textwidth]{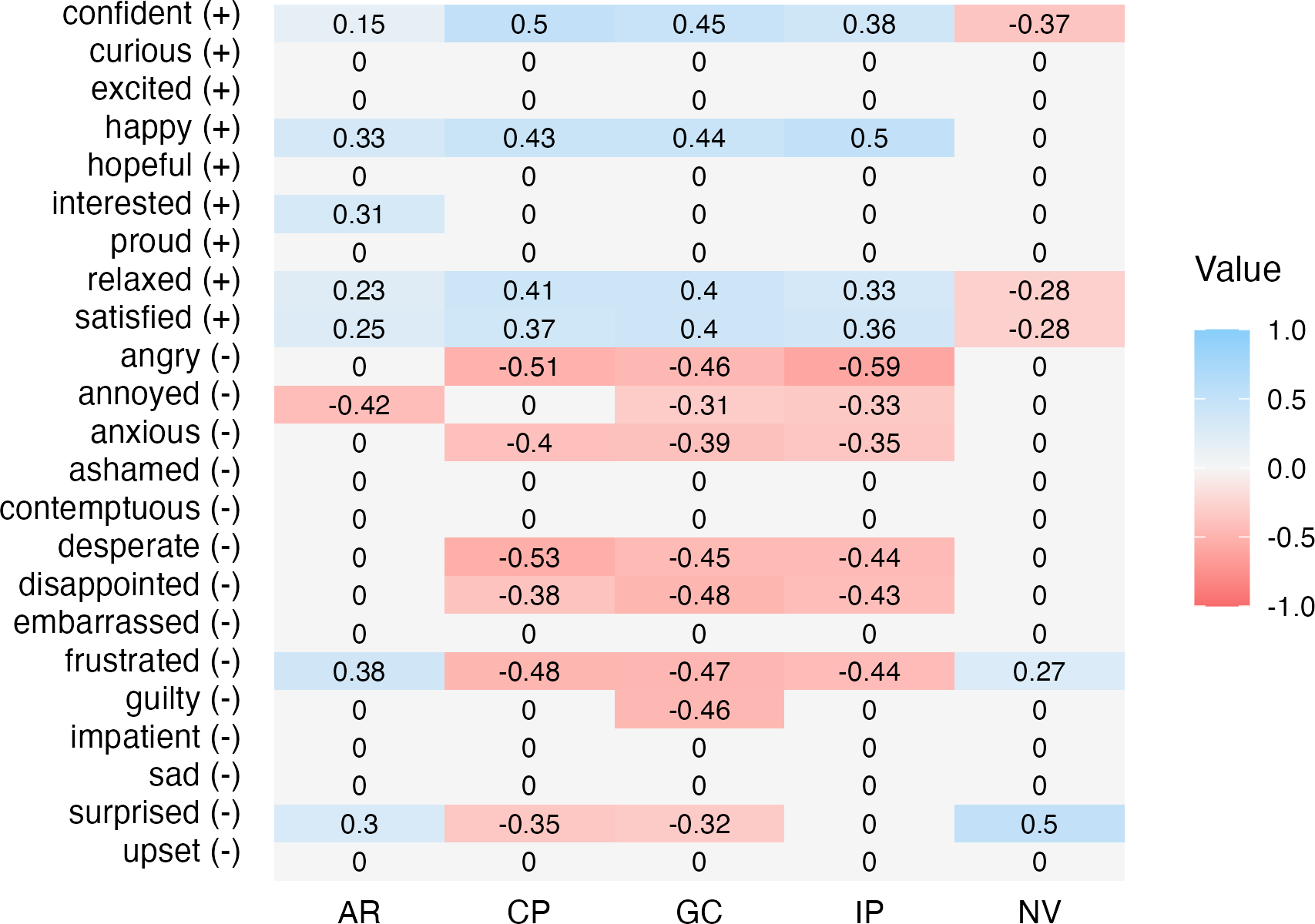}
\caption{Heatmap of the PCC between affective labels and summary values of appraisal dimensions aggregated across all annotators. The sign (+) or (-) after each label denotes whether it is positive or negative, respectively. Dimensions are as follows: AR: arousal, CP: coping, GC: goal conduciveness, IP: intrinsic pleasantness, NV: novelty. The correlations shown represent the average between annotators for whom the correlation was significant (\textit{p} < 0.05) after Bonferroni correction and of identical direction. Positive or negative direction is shown in blue or red, respectively. A correlation present for a single annotator was considered null.}

%\caption{Heatmap of the PCC between each dimension (summary value) presented in the columns -- AR: arousal, CP: coping, GC: goal conduciveness, IP: intrinsic pleasantness, and NV: novelty -- and each affective label on the rows, which are sorted in alphabetical order from positive (+) to negative (-). %, for the mean of the evaluations from the six annotators.
%Significant correlations (p-value $\leq{0.05}$) are coloured either in blue for positive values, or red for negative values. Non-significant correlations (p-value $>0.05$) are presented with 0 shown in grey. %Dark colours correspond to high correlation coefficient values between a given dimension and a given label, while light colours represent low correlation values.}
\label{fig_2}
\end{figure}

The arousal dimension was thought to be more difficult to annotate, as the context of the study, i.e. exercising on a computer and conversing with a virtual assistant, is likely to provide little variation for this dimension. This assumption is supported by the fact that the variance of arousal dimension was systematically among the lowest, which was observed for both time-continuous annotations -- arousal: .010, novelty: .022, intrinsic pleasantness: .009, goal conduciveness: .017, coping: .020 --, and summary annotations -- arousal: .008, novelty: .026, intrinsic pleasantness: .012, goal conduciveness: .021, coping: .027. 

%.0101, novelty : .0222, intrinsic pleasantness : .0094, goal conduciveness : .0168, coping : .0204) and for summary annotations (arousal : .00787, novelty : .0256, intrinsic pleasantness : .0118, goal conduciveness : .0207, coping : .0256). 

%On a per-session basis, the analysis of the agreement led to similar observations as the over all videos. This can be due to the fact that the videos may have presented similar scenarios and stimuli across different sessions, leading to consistent responses from the annotators.

%for continuous dimensions across all videos, all sessions, and on a per-video basis. However, since there was only one annotation per video in dimension summaries, it was not feasible to calculate the agreement per video. 

%On the other hand, when examining the agreement per video, we observed a decrease in the agreement for coping, goal conduciveness, and intrinsic pleasantness, highlighting their sensitivity to variations introduced by the video context. In contrast, arousal and novelty dimensions demonstrated a more consistent and less sensitive agreement per video. 
%Additionally, it is worth mentioning that the PCC values were higher than the CCC values, indicating the limitation of the PCC to consider some discrepancies or biases that may affect the overall agreement.

\subsection{Relation between dimensions and affective labels}
\label{sec:relation}

Further analysis of annotator agreement was conducted by exploring the relationship between summary values of appraisal dimensions and affective labels of each annotator separately. Specifically, the PCC was calculated between each pair of appraisal dimensions and affective labels for all sequences in which the value of the label concerned differed from zero; cf. section III.B of the supplementary material for a comprehensive view of the heatmap of each annotator. A correlation was considered consistent if it was statistically significant and in the same direction for at least two annotators. The average of consistent correlations between annotators are reported in Figure \ref{fig_2}. The labels that are consistently correlated with at least one appraisal dimension are marked with "\cmark" in the Table \ref{tab:coreset-overview}.

Broadly speaking, our results are in line with the predictions of appraisal theories regarding the link between affective labels and dimensions: positive affective labels are correlated with the ability to cope, goal conduciveness, and pleasantness, while negative affective labels are correlated with disability to cope, goal obstructiveness, and unpleasantness. 

Some of our results are, however, unexpected. The anger affective label is correlated with an absence of coping, whereas we would expect the opposite~\cite{SCHERER19-Dynamic}, as it generally refers to a situation in which an event is appraised as goal obstructive, but individuals appraise themselves capable of coping with it, e.g., being mugged in the street but wanting to defend themselves. In the context of CCT, where participants sit in front of a computer, different forms of anger may likely occur, probably involving an inability to cope. As expected, the affective label of surprise is highly correlated with novelty. Even though the surprise affective label is generally considered to be positive or negative depending on the context, it seems to be rather negative in our results as it is correlated with goal obstructiveness, and no coping. This could be explained by the fact that the occurrence of unexpected situations were more related to negative events such as unexpected negative feedback from the virtual assistant.

\begin{figure}[t!]
\includegraphics[width=0.47\textwidth]{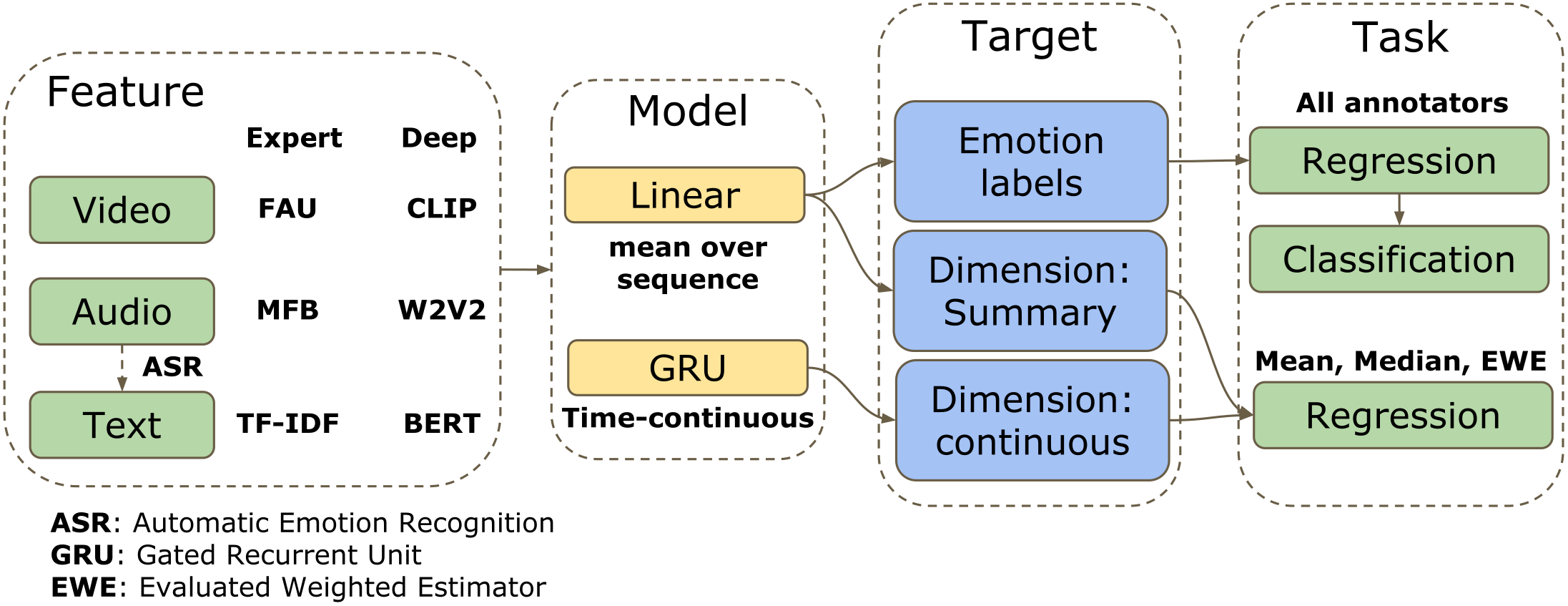}
\centering
\caption{Overview of the predictive experiments performed on the corpus. }
\label{fig:emo_rec_overview}
\end{figure}

\subsection{Label core set selection}
A core set of the likely most relevant affective labels in the context of AI-assisted CCT was defined based on our analyses. %Results of these analyses are summarised in the Table \ref{tab:coreset-overview}. 
Four main criteria of selection were used: (i) the frequency is significantly higher than the average on all affective labels; (ii) the agreement on presence is significantly higher than chance; (iii) the agreement on intensity is sufficiently high; and (iv) there is a significant correlation, in the same direction, with at least one dimension for at least two annotators. Ten affective labels meet these four criteria, five positive and five negative, cf. Table \ref{tab:coreset-overview}. %As a result of applying the above criteria, the core set of affects was carefully selected to represent frequently annotated labels with a high level of agreement between annotators. %This comprehensive approach ensured the reliability of the core set, leading to an accurate modelling of the affective labels described in the next section.

\section{Automatic recognition of affective states}
\label{sec: model}
In this section, we report on the automatic prediction of affective labels and dimensions. An overview of the experiments performed in this section is given in Figure \ref{fig:emo_rec_overview}. In brief, we extract different types of representations (expert or deep) from audio, textual, and visual data, and train specific models for the prediction of the affective labels and dimensions. 

All experiments were conducted using a single partitioning of the dataset, divided into three subsets: training, validation and test, comprising 56\,\%, 23\,\% and 21\,\% of the participants respectively; see Table \ref{tab:partition}. The partitions were carefully designed to minimise bias while maintaining key statistical properties, including: (i) the gender ratio (female vs. male); (ii) the proportion of healthy vs. MCI participants; (iii) the distribution of participants across affective induction conditions; (iv) the spectrum of education levels; and (v) the distribution of label presence, label intensity, and dimensional intensity. Heatmaps depicting these distributions are available in section IV of the supplementary material.

%In the following, we introduce the used representations of the audio, visual, and textual data, along with the predictive models and strategies, and then report on the results.

\begin{table}[t!]
\caption{Summary statistics of the partitions}
\label{tab:partition}
\setlength{\tabcolsep}{3pt}
\renewcommand{\arraystretch}{1.2}
\begin{tabular}{lrrrr}
\hline
\multicolumn{1}{c}{\textbf{}} & \multicolumn{1}{r}{\textbf{Training}} & \multicolumn{1}{r}{\textbf{Val}} & \multicolumn{1}{r}{\textbf{Test}} & \multicolumn{1}{r}{\textbf{Total}} \\ \hline
\textbf{Participants}& 34& 14& 13& 61\\
\textbf{Videos}& 1110& 851& 774& 2735\\
\textbf{Duration}& 2h44m36s& 1h51m42s& 1h44m28s& 6h20m47s\\ \hline

\multicolumn{5}{c}{\textbf{Population group}}\\ \hline
\textbf{Senior}& 29& 12& 11& 52\\
\textbf{MCI}& 5& 2& 2& 9\\ \hline

\multicolumn{5}{c}{\textbf{Gender (seniors)}}\\ \hline
\textbf{Male}& 6& 3& 3& 12\\
\textbf{Female}& 23& 9& 8& 40\\ \hline

\multicolumn{5}{c}{\textbf{Emotion Induction (seniors)}}\\ \hline
\textbf{With induction}& 18& 7& 7& 32\\ %\hline
\textbf{Without induction}& 11& 5& 4& 20\\ \hline

\multicolumn{5}{c}{\textbf{Education (seniors)}}\\ \hline
\textbf{High school}& 5& 2& 1& 8\\ 
\textbf{Professional certificate}& 3& 1& 2& 6\\ 
\textbf{Undergraduate}& 9& 5& 3& 17\\ 
\textbf{Graduate}& 12& 4& 5& 21\\ 
\hline

\end{tabular}
\end{table}

%\subsection{Data partitioning}
%\begin{enumerate}
    %\item Ratio of female vs. male participants
    %\item Ratio of senior vs MCI participants
    %\item Ratio of participants based on affective induction
    %\item Distribution of education levels of the senior participants
%\end{enumerate}

\subsection{Representations of audio, textual and video data}
\label{sec:rep}

We included both expert and deep representations of audio, visual, and textual\footnote{In order to make the comparison fair, textual data did not rely on the human transcriptions but on the outputs of an ASR, which is evaluated in detail in section IV.A of the supplementary material.} data to evaluate their respective contribution in the analysis of affect. Expert based features rely on the statistical description of perceptually salient patterns in the data, and involve relatively light-weight computational resources, whereas deep representations exploit self-supervised learning (SSL) methods on large scale datasets~\cite{Alisamir21-OTE}.%, and require more computational ressources.

\subsubsection{Expert based representations}
We used the Mel-scale Filter Banks (MFBs)~\cite{davis1980comparison}, Term Frequency - Inverse Document Frequency (TF-IDF)~\cite{salton1988term}, and Facial Action Units (FAUs)~\cite{ekman1978manual}, as representations of the audio, textual, and visual data, respectively. 

MFBs are popular features in acoustic processing. They are based on a simple signal processing pipeline which consists in taking the Fourier transform of the acoustic signal through a sliding window (25\,ms shifted forward in time every 10\,ms), and then mapping the power spectrum to a mel-scale (80 filter banks), which is a non-linear perceptual scale based on the human auditory system. MFBs were extracted using the SpeechBrain toolkit~\cite{speechbrain}.%, and standardised on the training partition. 

The TF-IDF is a classical feature in text processing, capturing the frequency of a word within a document (TF), relative to its rarity across a given set of documents (IDF). The method therefore tokenises the words of a sentence, which is represented by a vector of the count of the words it contains, and further normalised by the term frequency across the entire training partition of the dataset. %Thus, the TF-IDF provides a vector of the size of the vocabulary, and the obtained textual representation does not encode any semantic information.
%, which is 1173 words on the ASR transcriptions of the training partition of the THERADIA WoZ corpus
%On the other hand, because of its simplicity, it can be computed very quickly and at low cost, compared to more complex methods that take semantic or contextual information into account. 

FAUs describe the intensity of movements of facial muscles related to an apparent facial expression, and are commonly used as facial descriptors in affective computing. 
%Each action unit represents a specific action of a muscle, or group of muscles, and provide information about the activation or intensity of that specific facial movement. 
FAUs were extracted using OpenFace~\cite{openface}, which provides estimations of the intensity of a set of 17 FAUs with a framerate of 30\,Hz.%: `AU01-inner brow raiser', `AU02-outer brow raiser', `AU04-brow lowerer', `AU05-upper lid raiser', `AU06-cheek raiser', `AU07-lig tightener', `AU09-nose wrinkler', `AU10-upper lid raiser', `AU12-lip corner puller', `AU14-dimpler', `AU15-lip corner depressor', `AU17-chin raiser', `AU20-lip stretcher', `AU23-lip tightener', `AU25-lips part', `AU26-jaw drop', and `AU45-blink'. 

\subsubsection{Self-supervised representations}
%A major benefit of SSL approaches is their ability to leverage the ever-growing mass of available unlabeled data to model patterns of the encoded high-level information~\cite{PARCOLLET2024101622}. 
%SSL strategies are various and may be divided, for speech, into four families: i) generative -- the input signal is reconstructed after various corruptions, e.g., Mockingjay~\cite{liu2020mockingjay}, Tera~\cite{liu2021tera}, or data2vec~\cite{baevski2022efficient}, ii) predictive -- discrete labels are predicted after a clustering of the input samples, e.g., WavLM~\cite{chen2021wavlm}, HuBERT~\cite{hsu2021hubert}, or BEST-RQ~\cite{chiu2022self}, iii) contrastive -- positive and negative candidates originating from the given signal are distinguished, e.g., Wav2Vec 2.0 (W2V2)~\cite{baevski2020wav2vec} or Contrastive Predicting Coding (CPC)~\cite{oord2018representation}, and iv) multi-task -- different objectives or modalities are combined to build a rich feature extractor, e.g., PASE+~\cite{ravanelli2020multi}. 
Deep representations of audio, textual, and video data include the following models: Wav2Vec2~\cite{baevski2020wav2vec}, BERT~\cite{devlin2018bert}, and CLIP~\cite{clip}, respectively.

We used a multilingual Wav2Vec2 model that exploits both convolutional and transformer layers to distinguish positive and negative samples of speech in a contrastive approach. The model\footnote{\url{https://huggingface.co/voidful/wav2vec2-xlsr-multilingual-56}} we used has a large architecture, and was trained on 56 different languages~\cite{conneau2020unsupervised}, including French, and then fine-tuned for ASR. % and the corresponding audio embeddings have a size of 1024 features.

Transformer-based representations such as BERT are popular in text processing, as they can encode the context of the words, which is crucial for affect related tasks~\cite{sun2019utilizing, yang2019emotionx}. %Moreover, BERT is pre-trained in a self-supervised manner on huge amounts of text, which makes it possible to provide effective textual representations without the use of labelled data. 
We used a BERT model\footnote{\url{https://huggingface.co/nlptown/bert-base-multilingual-uncased-sentiment}} fine-tuned for sentiment analysis on six languages (English, Dutch, German, French, Spanish, and Italian), and with the base architecture.% and provides embeddings with a size of 768. 

CLIP\footnote{\url{https://openai.com/research/clip}} is a model developed by OpenAI~\cite{clip} and trained on a large dataset that includes both images and their corresponding textual descriptions allowing the model to learn the semantic relationship between visual and textual information. The architecture consists of image and text encoders that encode the input image and text into fixed-length embeddings using a contrastive learning mechanism.%, where the shared embedding space allows for a semantic understanding of images.
%This constrastive learning ensure that positive image-text pairs are closer together in a shared embedding space, while negative pairs are pushed farther apart. 
 %In this study, the pre-trained CLIP model\footnote{\url{https://openai.com/research/clip}} provides embeddings with a size of 512.

\subsection{Models and training strategies}
\label{sec:models}

The aforementioned multimodal representations were used to model the affective labels and dimensions. We evaluated the three following predictive models for the labels intensity and dimensions summaries:

\begin{enumerate}
    \item Linear: One linear layer on top of the representations to map the features to the desired number of outputs.
    \item Multi-Layer Perceptron (MLP): The linear layer as above, combined with a hidden linear layer whose size was the number of features divided by two.
    \item Gated Recurrent Units (GRU): A GRU model with the hidden size defined to be the feature size divided by two.
\end{enumerate} 
Outputs of the Linear and MLP models were averaged over the sequence, whereas the last output of the GRU model was used to represent the output of the whole sequence. 
Preliminary results obtained in the prediction of the affective labels showed that the performance of the MLP model was consistently better than the performance of Linear and GRU models, both for the classification and regression tasks.%, which might be due to the relatively short amount of context available in the sequences. The MLP model was therefore consistently used in our experiments for the prediction of the affective labels. %Furthermore, using only linear layers on top of self-supervised representations is nowadays the state-of-the-art in emotion recognition~\cite{evain:hal-03407172, siriwardhana2020jointly}.

For dimension summaries, the model was trained to predict the five dimensions altogether in a multi-task approach. Our first experiments revealed that the Linear model consistently outperformed both the MLP and GRU models. %, which could be attributed to the specific nature of the task, where the Linear model is possibly better capturing the dependencies between the dimensions by simplifying the prediction process.

To predict the time-continuous annotations of the dimensions, it was necessary to resample certain features to align with the annotation frame rate of 10\,Hz. Audio representations, sampled at 100\,Hz for MFBs, and 50\,Hz for Wav2vec2, were averaged into 10\,ms chunks to match the annotation frequency. The same strategy was applied to the FAUs and CLIP visual representations. For textual representations, we utilised BERT exclusively, as TF-IDF provides a single feature vector for the entire sequence. Textual features were aligned to the sequence length using spline interpolation, ensuring a smooth and continuous representation of the features over time.
%, and was compared with other interpolation techniques, including linear interpolation, but found that spline interpolation achieved better results. However, for TF-IDF features, alignment was not applicable because they are based on word frequency at the sentence level.% and produce a feature vector of the same size as the vocabulary. %Therefore, any adjustment will distort the original representation.
Since the Linear and MLP models did not perform well in predicting time-continuous dimensions, we explored various GRU model architectures:

\begin{enumerate}
     \item One-Layer GRU: A single-layered GRU with a hidden size set to either half the input feature size or equal to the input feature size.
    \item Two-Layers GRU: Similar to the One-Layer GRU but with two hidden layers.
    \item Three-Layers GRU: Similar to before, but with three hidden layers. 
\end{enumerate}
Additionally, we tested a fixed set of hidden sizes across all GRU models: 128, 256, and 512. The results indicated that a GRU model with three hidden layers, each matching the input size, performed optimally for deep representations. In contrast, the model with a hidden size of 256 produced the best results for hand-crafted representations.%, which might be explained by the complexity of the task, where the model needs to capture temporal patterns in the data.

For multimodal fusion, we explored both middle and late fusion techniques to address the issue of asynchronous inputs. The best performing approach for both labels and dimension summaries was decision-based fusion using an Ordinary Least Squares (OLS) regression model\footnote{\url{https://scikit-learn.org}}. %, which also presents the advantage to quantify the contribution of the modalities in the decision by returning coefficients. 
For time-continuous dimensions, fusion was achieved through a GRU functioning as a weighted averaging layer, effectively capturing the temporal dynamics of the features. The GRU weights were optimised on the training set. %for each modality to take into account their respective temporal dynamics. 
%that was t learned during the training process.% and quantify the contributions of the modalities in the prediction of the time-continuous dimensions. 

All the models were trained with the Adam optimizer and an initial learning rate of $10^{-3}$ for affective labels and $10^{-4}$ for dimensions. The batch size and gradient accumulation were set to 10 for labels and 1 for dimensions. The maximum number of training epochs was set to 50 and an early stopping strategy was used with five epochs. % if the loss on the validation set did not decrease after five epochs. %For time-continuous dimensions prediction, the GRU models consisted of three layers, and the size of the hidden layers was equal to the size of the input features.
% add details abt features used in continuous case
%grid search : nb hidden layers (1-4), size of the hidden layers (256 - 512 - 768 - 1024)
%input size FAU: 17, CLIP: 512, W2V2: 1024, MFB: 80, TF-IDF: 1129, BERT: 768
The models were trained using the Mean Squared Error (MSE) as the loss function, and for evaluation, the Concordance Correlation Coefficient (CCC) was calculated on all sequences~\cite{Weninger16-DTR}. All experiments were carried out using Pytorch\footnote{\url{https://pytorch.org}}\cite{paszke2019pytorch}, with random seeds set to zero. The computer's operating system was Debian GNU/Linux 10, and the GPU used to train the models was an NVIDIA Quadro RTX 6000 with 23 gigabytes of memory, CUDA version 11.3.

% Moreover, to train the models described above based on the text modality, we used automatic transcriptions. The rationale and the analysis of this choice is further explained below.

\subsection{Affect prediction for the core set labels}
\label{sec:affectIntensityPred}

%Experiments involved training a model to predict all the six available annotations for each affective label. However, to evaluate each trained model, the average of the predictions is compared with the average of the six label targets, similar to~\cite{Ringeval15-POA}. %Therefore, for each affective label, a model is trained to predict six outputs (for the six annotators). 

The results for label intensity predictions, presented in Table \ref{tab:label-regression}, reveal some strong dependencies between modalities and affective labels; see section IV.B of the supplementary material for additional details on prediction bias related to population demographics. For instance, the label ``happy'' was most accurately assessed via the video modality, ``interested'' through the textual modality, and ``frustrated'' and ``surprised'' using the audio modality with deep representations. The choice between expert-crafted and deep features also introduced notable differences in the results according to the modality. For example, the label ``frustrated'' achieved a significantly higher CCC (.327) with TF-IDF features compared to MFB (.151), whereas the reverse trend was observed when using SSL representations. Overall, deep representations across textual, audio, and video modalities consistently outperformed their hand-crafted counterparts on average.

The fusion of the different modalities yielded the best results on average and for most affective labels. An analysis of the fusion weights revealed that the importance of each modality varied significantly based on its representation. For instance, when using expert features, TF-IDF had a substantially greater influence on the results, with an averaged fusion coefficient of 0.8, compared to 0.1 for audio (MFBs) and video (FAUs). However, with deep representations, the contributions of audio and video surpassed that of text, with average coefficients of .28, .35, and .35 for text, audio, and video respectively.

\begin{table}[t!]
\setlength{\tabcolsep}{4.5pt}
\scriptsize
%\tiny, \scriptsize, \footnotesize, \small, \normalsize, \large, \Large, \LARGE, \huge, \Huge
  \caption{Core-set label intensity prediction results (CCC)}
  \label{tab:label-regression}
  \centering
  
  \begin{tabulary}{1.0\textwidth}{m{0.05\textwidth}|R{0.03\textwidth}R{0.03\textwidth}R{0.03\textwidth}R{0.03\textwidth}|R{0.03\textwidth}R{0.03\textwidth}R{0.03\textwidth}R{0.03\textwidth}}

\hline
  \multirow{3}{*}{Label} & \multicolumn{4}{c|}{Expert Representations} & \multicolumn{4}{c}{Deep Representations} \\ \cline{2-9}
 & \multicolumn{1}{c}{Text} & \multicolumn{1}{c}{Audio} & \multicolumn{1}{c}{Video} & \multicolumn{1}{c|}{\multirow{2}{*}{Multi.}} & \multicolumn{1}{c}{Text} & \multicolumn{1}{c}{Audio} & \multicolumn{1}{c}{Video} & \multicolumn{1}{c}{\multirow{2}{*}{Multi.}} \\ %\cline{2-4} \cline{6-8}
 & \multicolumn{1}{c}{TF-IDF} & \multicolumn{1}{c}{MFB} & \multicolumn{1}{c}{FAU} & \multicolumn{1}{c|}{} & \multicolumn{1}{c}{BERT} & \multicolumn{1}{c}{W2V2} & \multicolumn{1}{c}{CLIP} & \multicolumn{1}{c}{} \\ \cline{1-9}
Annoyed & .075 & .047 & .274 & .255 & .255 & .231 & .311 & .280 \\
Anxious & .115 & .081 & .129 & .171 & .087 & .192 & .123 & .214 \\
Confident & .233 & .176 & .168 & .422 & .473 & .482 & .247 & .487 \\
Desperate & .172 & .148 & .193 & .259 & .221 & .277 & .175 & .252 \\
Frustrated & .327 & .151 & .175 & .387 & .222 & .340 & .239 & .465 \\
Happy & .117 & .273 & .446 & .490 & .108 & .303 & .462 & .508 \\
Interested & .147 & .135 & -.023 & .207 & .364 & .204 & .133 & .300 \\
Relaxed & .314 & .300 & .385 & .515 & .363 & .245 & .440 & .529 \\
Satisfied & .323 & .225 & .253 & .428 & .057 & .287 & .326 & .486 \\
Surprised & .167 & .165 & .024 & .197 & .143 & .238 & .095 & .275 \\ 
\hline
Average & .199 & .170 & .202 & .333 & .229 & .280 & .255 & .380 \\ 
\hline

  \end{tabulary}
  
\end{table}

An affective label was considered present in a given sequence if at least one annotator reported an intensity value for that label. To predict these labels, we explored two modeling approaches: training a binary classification model for each affective label or deriving a threshold from the regression models. Since the binary classification approach yielded significantly poorer results compared to the thresholded-based decision from regression models, we opted for the latter in our experiments.

%Also, the threshold was optimised on the validation set for each affect label from a range of zero to 100 (same as intensity), with intervals of five. 

The results, presented in Table \ref{tab:label-classification}, indicate some dependencies between the best-performing modality and the affective labels, although these differences are less pronounced than in previous experiments. Overall, no significant performance gap was observed between expert and deep representations across modalities, with the exception of the audio modality. Additionally, fusion consistently enhanced performance across most scenarios.

\subsection{Dimensions summaries prediction}
\label{dim_sum_Sec}

\begin{table}[t!]
\setlength{\tabcolsep}{4.5pt}
\scriptsize
%\tiny, \scriptsize, \footnotesize, \small, \normalsize, \large, \Large, \LARGE, \huge, \Huge
  \caption{Core-set label classification results (\%UAR)}%Note that the textual results are based on automatic transcriptions.
  \label{tab:label-classification}
  \centering
  
  \begin{tabulary}{1.0\textwidth}{m{0.05\textwidth}|R{0.04\textwidth}R{0.03\textwidth}R{0.03\textwidth}R{0.03\textwidth}|R{0.03\textwidth}R{0.03\textwidth}R{0.03\textwidth}R{0.03\textwidth}}

\hline
  \multirow{3}{*}{Label} & \multicolumn{4}{c|}{Expert Representations} & \multicolumn{4}{c}{Deep Representations} \\ \cline{2-9}
 & \multicolumn{1}{c}{Text} & \multicolumn{1}{c}{Audio} & \multicolumn{1}{c}{Video} & \multicolumn{1}{c|}{\multirow{2}{*}{Multi.}} & \multicolumn{1}{c}{Text} & \multicolumn{1}{c}{Audio} & \multicolumn{1}{c}{Video} & \multicolumn{1}{c}{\multirow{2}{*}{Multi.}} \\ %\cline{2-4} \cline{6-8}
 & \multicolumn{1}{c}{TF-IDF} & \multicolumn{1}{c}{MFB} & \multicolumn{1}{c}{FAU} & \multicolumn{1}{c|}{} & \multicolumn{1}{c}{BERT} & \multicolumn{1}{c}{W2V2} & \multicolumn{1}{c}{CLIP} & \multicolumn{1}{c}{} \\ \cline{1-9}
 
Annoyed & 63.2 & 57.5 & 62.4 & 66.0 & 63.1 & 63.7 & 65.1 & 69.4 \\
Anxious & 66.3 & 59.3 & 57.0 & 67.2 & 65.5 & 64.9 & 58.2 & 67.6 \\
Confident & 70.4 & 66.3 & 59.8 & 71.6 & 73.4 & 73.1 & 59.5 & 76.0 \\
Desperate & 71.3 & 68.6 & 62.4 & 72.1 & 69.3 & 74.9 & 65.6 & 75.7 \\
Frustrated & 72.2 & 63.9 & 63.3 & 74.1 & 70.4 & 69.9 & 63.4 & 74.5 \\
Happy & 65.4 & 62.5 & 71.1 & 73.5 & 65.8 & 66.8 & 73.9 & 76.7 \\
Interested & 61.7 & 57.5 & 50.1 & 62.4 & 64.9 & 60.5 & 59.1 & 64.1 \\
Relaxed & 68.7 & 64.9 & 68.7 & 75.1 & 69.0 & 67.3 & 64.1 & 74.0 \\
Satisfied & 70.4 & 63.9 & 64.0 & 72.2 & 73.9 & 70.0 & 63.8 & 75.3 \\
Surprised & 62.3 & 56.8 & 52.2 & 61.8 & 62.1 & 61.7 & 54.1 & 62.3 \\
\hline
Average & 67.2 & 62.1 & 61.1 & 69.6 & 67.7 & 67.3 & 62.7 & 71.6\\
\hline

  \end{tabulary}
\end{table}

%Annoyed & 63.2\% & 57.5\% & 62.4\% & 66.0\% & 63.1\% & 63.7\% & 65.1\% & 69.4\% \\
%Anxious & 66.3\% & 59.3\% & 57.0\% & 67.2\% & 65.5\% & 64.9\% & 58.2\% & 67.6\% \\
%Confident & 70.4\% & 66.3\% & 59.8\% & 71.6\% & 73.4\% & 73.1\% & 59.5\% & 76.0\% \\
%Desperate & 71.3\% & 68.6\% & 62.4\% & 72.1\% & 69.3\% & 74.9\% & 65.6\% & 75.7\% \\
%Frustrated & 72.2\% & 63.9\% & 63.3\% & 74.1\% & 70.4\% & 69.9\% & 63.4\% & 74.5\% \\
%Happy & 65.4\% & 62.5\% & 71.1\% & 73.5\% & 65.8\% & 66.8\% & 73.9\% & 76.7\% \\
%Interested & 61.7\% & 57.5\% & 50.1\% & 62.4\% & 64.9\% & 60.5\% & 59.1\% & 64.1\% \\
%Relaxed & 68.7\% & 64.9\% & 68.7\% & 75.1\% & 69.0\% & 67.3\% & 64.1\% & 74.0\% \\
%Satisfied & 70.4\% & 63.9\% & 64.0\% & 72.2\% & 73.9\% & 70.0\% & 63.8\% & 75.3\% \\
%Surprised & 62.3\% & 56.8\% & 52.2\% & 61.8\% & 62.1\% & 61.7\% & 54.1\% & 62.3\% \\
%\hline
%Average & 67.2\% & 62.1\% & 61.1\% & 69.6\% & 67.7\% & 67.3\% & 62.7\% & 71.6\%\\

Results are presented in table \ref{dim_s} for two different gold standards: the average of annotations and the median; see section IV.B of the supplementary material for further details on prediction bias related to population demographics. Comparing these gold standards, the average of annotations consistently yielded the best performance across all five dimensions. In terms of individual dimensions, arousal was best predicted by the audio modality, with no significant improvement observed from fusion with the other modalities. It also had the lowest scores among the dimensions, which might be explained by the factors discussed in Section \ref{sec:relation}. 
Coping was the most accurately predicted dimension, with all modalities performing similarly well and demonstrating complementary effects during fusion. Goal conduciveness was best predicted by textual representations, with fusion incorporating audio and video modalities further improving performance. Intrinsic pleasantness was most effectively captured by the video modality, with textual representations proving relevant only when computed over the entire sequence using TF-IDF. Deep audio representations also performed comparably well. Finaly, novelty was particularly challenging for the video modality, which yielded extremely low scores, while textual representations proved to be more effective in capturing  this dimension.

By analysing the relative importance of each modality based on the coefficients from the linear regression model, we found that audio features contributed very little to predicting arousal, especially when using expert features. This could explain the drop in performance observed with the multimodal approach, compared to using the audio modality alone. For the others dimensions, the contributions of each modality were influenced by the complexity of the representations. When using deep features, the audio modality emerged as the most influential. However, this contribution was diminished when expert features were used, with textual information playing a more significant role.

\begin{table}[t!]
\caption{Dimension summary prediction (CCC)}
\label{dim_s}
\setlength{\tabcolsep}{1.5pt}
\renewcommand{\arraystretch}{1.8}
\scriptsize

\begin{tabular}{cccccccccc}
\hline
& \multicolumn{9}{c}{Modalities}\\ \hline
& & \multicolumn{2}{c}{Text} & \multicolumn{2}{c}{Audio}             & \multicolumn{2}{c}{Video} & \multicolumn{2}{c}{Multimodal}           \\ \hline
Dimensions & GS & TF-IDF & BERT      & MFB & W2V2 & FAU& CLIP& Expert & Deep \\ \hline

\multirow{2}{*}{Arousal} 
& Mean   & .311 & .195 & .390 & .309 & .234 & .209 & .356  & .310 \\ \cline{2-10} 
& Median & .264 & .178 & .276& .282 & .252 & .219 &  .326  & .276 \\ \hline

\multirow{2}{*}{Coping}
& Mean   & .506 & .452 & .294 & .499 & .336 & .460 & .599  & .684 \\ \cline{2-10} 
& Median & .488 & .459 & .302& .493 & .329 & .497  &  .603 &  .666 \\ \hline

\multirow{2}{*}{Goal conduciveness}    & Mean   & .520 & .495      & .347 & .428 & .345 & .380 & .630  & .662\\ \cline{2-10} 
& Median & .491 & .468 & .291& .426 & .361 & .389 &   .605 &  .642\\ \hline

\multirow{2}{*}{Intrinsic pleasantness} & Mean   & .399 & .313 & .296 & .357 & .500 & .573 &  .606  & .609 \\ \cline{2-10} 
& Median & .376 & .202 & .261& .346 & .483 & .463 &  .574 & .594 \\ \hline

\multirow{2}{*}{Novelty}
& Mean   & .429 & .377 & .357& .371 & .015 & .312 & .465  &  .559 \\ \cline{2-10} 
& Median & .329 & .265      & .277& .298 & .021 & .172 & .376  & .420 \\ \cline{2-10}
\hline
\end{tabular}
\end{table}

\subsection{Continuous dimensions prediction}

%In this section, we conducted experiments to predict the five dimensions --coping, goal conduciveness, intrinsic pleasantness, arousal, and novelty-- at each time stamp of 100 ms for each sequence (the same as the sampling rate of annotations). As described in Section \ref{sec:models}, a GRU model was trained with five outputs, each corresponding to one of the dimensions using the CCC as the loss function. 
Results are presented in Table \ref{dim_c} for three gold standards: the average and the median of the time-continous annotations, and a version based on the Evaluator Weighted Estimator (EWE), which calculates a weighted mean of the annotations using the average cross-CCC between each rater's scores as weights~\cite{stappen2021muse}; for further details on prediction bias based on population demographics, see section IV.B of the supplementary material. Overall, the EWE approach consistently outperformed the others two gold standards across the five dimensions.
%Table \ref{dim_c} summarises the obtained performance in terms of CCC for the three modalities using traditional and deep features. We reported the results trained for the following \textit{gold standards}: 1) the mean of annotations, 2) the median, and 3) the Evaluator Weighted Estimator (EWE). EWE is a weighted mean of annotations, based on the cross-correlations of CCCs between each rater's annotation and the mean of all other annotations \cite{stappen2021muse}. The reason why we chose EWE was because it assigns weights to different annotators based on the CCC metric, which is suitable for continuous annotations, due to its ability to consider both the difference of means and the variances of annotations.
%The results of the prediction of continuous dimensions are consistent with the predictions of the dimension summaries in Table \ref{dim_s} discussed above. 

Arousal was the best predicted dimension, but this was primarily due to the low variance observed within this dimension, as explained in the previous section. For coping, goal conduciveness and novelty, Wav2vec2 audio representations outperformed textual features, which contrasts with the findings on the summarised dimensions. This discrepancy may be attributed to the fact that textual features provide a static representation, which may not capture the continuous variations within the data. Regarding intrinsic pleasantness, the three modalities performed similarly and proved complementary during fusion, which is consistent with the previous section's findings. % For intrinsic pleasantness, better performance was obtained using Wav2vec2 based features.
Additionally, the multimodal approach consistently improved performance. An analysis of the fusion weights revealed that when using the mean of annotations as the gold standard, the model evenly distributed importance across all modalities. However, when using the median or EWE gold standards, the audio modality emerged as the most significant, while text and video contributed similarly.

An examination of the weights learned by the GRU during fusion showed that when using the mean of annotations as the gold standard, the model assigned equal importance to all modalities. In contrast, with the  median and EWE gold standards, the audio modality became the most influential, while text and video modalities contributed similarly to the overall predictions. 

\begin{table}[t!]
\centering
\caption{Time-continuous dimension prediction (CCC)}
\label{dim_c}
\setlength{\tabcolsep}{3pt}
\renewcommand{\arraystretch}{1.5}
\scriptsize

\begin{tabular}{cccccccc}
\hline
&& \multicolumn{6}{c}{Modalities}\\ \hline
&& Text & \multicolumn{2}{c}{Audio} & \multicolumn{2}{c}{Video} & Multimodal \\ \hline
Dimensions& GS& BERT & MFB& W2V2 & FAU & CLIP & Deep \\ \hline

\multirow{3}{*}{Arousal}& Mean   & .533 & .575& .545 & .539& .500& .545\\ \cline{2-8} 
& Median & .644 & .705& .665 & .696& .643& .684\\ \cline{2-8} 
& EWE & .554 & .709 & .699 & .651 & .643 & .691\\ \hline

\multirow{3}{*}{Coping} & Mean   & .356 & .254& .281 & .198& .239& .346 \\ \cline{2-8} 
& Median & .186 & .269 & .406 & .213 & .260 & .340 \\ \cline{2-8} 
& EWE    & .376 & .308& .422 & .183& .251& .418 \\ \hline

\multirow{3}{*}{Goal conduciveness}     & Mean   & .320 & .230& .236 & .207& .199& .357 \\ \cline{2-8} 
& Median & .168 & .224& .390 & .218& .249& .347\\ \cline{2-8} 
& EWE    & .310 & .248 & .339 & .191& .183& .420 \\ \hline

\multirow{3}{*}{Intrinsic pleasantness} & Mean   & .258 & .202& .188 & .262  & .231 & .317\\ \cline{2-8} 

& Median & .125 & .195 & .308 & .272& .256& .324 \\ \cline{2-8} 
& EWE    & .262 & .250 & .289 & .242& .245 & .370 \\ \hline

\multirow{3}{*}{Novelty}& Mean   & .224 & .154& .196 & .094 & .175& .231 \\ \cline{2-8} 
& Median & .139  & .116& .308 & .111& .203& .274\\ \cline{2-8} 
& EWE    & .262 & .203& .279 & .071& .202& .339 \\ \hline
\end{tabular}
\end{table}

\section{Conclusion} % Sina: removed "and discussion" since we did not discuss anything in the end
\label{sec: conclusion}
This paper presents the THERADIA WoZ corpus, a valuable multimodal dataset for affective computing research in healthcare. Based on current affective science and appraisal theories, the corpus includes 39.5 hours of multimodal data from 52 healthy older participants and 9 with MCI performing WoZ-assisted cognitive training. To capture affective episodes, affect inductions were applied to 32 healthy older participants. The data is fully transcribed and partially annotated across four appraisal dimensions -- novelty, intrinsic pleasantness, goal conduciveness, and coping -- as well as arousal, and with 23 achievement-related affect labels. These dimensions, provided in both in continuous time and summarised forms, highlights ten core affective labels of significant relevance for AI-assisted CCT applications.

In-depth baseline experiments were conducted to model affect using text, audio, and video modalities, including investigations of multimodal fusion through both computationally efficient feature and contextual deep learning representations. Results revealed key dependencies between modalities and affect representations. This work provides new insights into the field of affect recognition, offering novel insights into affect recognition in healthcare.

%We further showed that the trained models generalised well to MCI participants, despite the fact that senior participants made up the majority of the training set. 

% Baseline experiments were carried out to assess the relationship between Action Units, word use frequency, acoustic features with dimensions in both continuous and summary values and summary values of label presence and intensity. In addition, a coreset of the 10 most relevant labels in the context of WoZ assisted CCT was identified. This work provides new insights to the field of affect recognition in healthcare and contributes to bring supplemental data for benchmark comparison.

This work contributes to the ongoing development of affective computing in interaction -- and specifically for healthcare -- providing a valuable dataset for benchmark comparisons. We believe that combining the strengths of affective computing and psychology, as demonstrated in this multidisciplinary study, will significantly advance our understanding of affect and its modeling, benefiting both fields and their applications.
\section{Acknowledgments}
This research has received funding from the Banque Publique d’Investissement (BPI) under grant agreement THERADIA, the Association Nationale de la Recherche et de la Technologie (ANRT), under grant agreement No. 2019/0729, and has been partially supported by MIAI@Grenoble-Alpes, (ANR-19-P3IA-0003).

\balance

\bibliographystyle{IEEEtran}
\bibliography{refs.bib}

    

\section*{Supplementary material}

# THERADIA WoZ: An Ecological Corpus for Appraisal based Affect Research in Healthcare

## Supplementary material

The supplementary material provides details for specific sub-sections of the main manuscript. The section and sub-section names in this document correspond to those in the main manuscript to aid in understanding which parts are being referred to, excepted for the last section which specifically deals with bias and fairness in the affect predictions.

## I. Data collection

### A. CCT supported by a Virtual Assistant

All participants carried out a CCT session, which was based on eight exercises selected from the HappyNeuronPro CCT platform[1]. The Figure 1 illustrates the on-screen trial of these exercises that aimed at training various aspects of memory (e.g., visual, verbal), planning, reasoning, mental arithmetic, and grammatical logic.

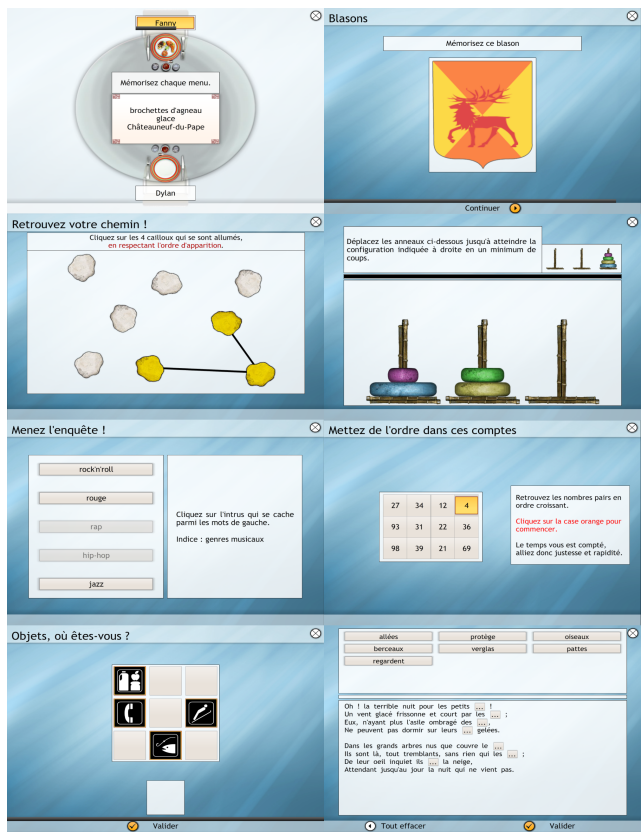

Fig. 1: Display of on-screen trial of the eight exercises selected for the data collection.

[1] https://www.happyneuron.fr

### B. Procedure

Induction was carried out for every second exercise in the induction groups. Three parameters were manipulated for induction: the exercise introduction, in which the difficulty was advertised as easy or hard regardless of the real difficulty of the exercise; the real difficulty of the exercise, in which the difficulty was easy or hard regardless of the advertised difficulty; and the feedback of the performance, in which participants were either a bit congratulated, highly congratulated, a bit criticised, or highly criticised, regardless of the performance. Induction groups were assigned to possible induction combinations on the basis of a Latin square, cf. Table I.

TABLE I: Types of induction assigned to the different groups according to exercise introduction, difficulty and feedback.

| Participant group | Task number | Exercise introduction induction | Real difficulty of the exercise | Feedback induction |
|---|---|---|---|---|
| A | 2 | Easy | Easy | Congratulation |
| A | 4 | Easy | Hard | Critical + |
| A | 6 | Hard | Easy | Critical |
| A | 8 | Hard | Hard | Congratulation + |
| B | 2 | Easy | Easy | Congratulation + |
| B | 4 | Easy | Hard | Congratulation |
| B | 6 | Hard | Easy | Critical + |
| B | 8 | Hard | Hard | Critical |
| C | 2 | Easy | Easy | Critical |
| C | 4 | Easy | Hard | Congratulation + |
| C | 6 | Hard | Easy | Congratulation |
| C | 8 | Hard | Hard | Critical + |
| D | 2 | Easy | Easy | Critical + |
| D | 4 | Easy | Hard | Critical |
| D | 6 | Hard | Easy | Congratulation + |
| D | 8 | Hard | Hard | Congratulation |

## II. Data segmentation, annotation, and anonymisation

### A. Segmentation, transcription, and annotation of sequences

The reference dictionary was "Le Petit Robert", and unconventional words were preceded with the commercial symbol. Punctuation was not transcribed, excepted for the question mark. The use of quotes, uppercase letters (except for proper nouns), and apostrophes was also restricted. Elisions were retained if the schwa maintained a melodic content without being assimilated to "euh". Truncated words were marked with a hyphen. Verbal expressions were preserved, with deformations within the lexicon retained and others corrected. Incorrect elisions were not corrected orthographically. Onomatopoeia and interjections were listed upon their spellings. The text further introduced diacritics for marking the beginning and end of virtual assistant interactions, as well as the presence of non-verbal information.

## B. Data annonymisation

In the public version of the dataset that will be released, the data has been anonymised to protect participants privacy by removing personal identifiers including participants names, ages, cities, family backgrounds, and occupations. We used regular expressions to detect these identifiers from sequences transcriptions. Name and location patterns were detected based on capitalised words, age patterns look for numerical numbers followed by "ans", and family and job patterns search for a list of keywords that might indicate someone's job, such as "travail", "profession", "retraité", or words related to family relationships, such as "fils", "fille", or "enfants". Any audio segments that contained private information were carefully removed, and corresponding intervals in the complete audio files were replaced by silence. Similarly, for the complete video file, frames containing private information have been replaced with a black screen.

## C. Data annotation

To ease the time-continuous annotation process, the position of the mouse was freeze once the video was played, and movements of the mouse were only applied to the annotation cursor which was always located at the middle of the scale. At the end of the time-continuous annotation, a graph showing the dynamics of the dimension over time was presented to the annotators, who could then validate or perform the annotation again as shown in Figure 2. After the validation, annotators were asked to provide a summary value for the whole sequence as depicted in the Figure 3. Finally, once all dimensions were annotated, annotators were asked to indicate if one or more labels could describe the affect they had recognised in the sequence and provide an intensity value for them as illustrated in the Figure 4. Labels were presented in different parts of the screen according to their valence. The order of the positive labels, and negative labels, were defined randomly for each annotator. In order to reduce the number of operations to be effected by the annotators to provide their rating, the annotation bar was displayed when a click was maintained on an annotation label, and vertical displacements of the mouse allowed to change the intensity value.

## III. CORPUS ANALYSIS

### A. Dimensional annotations

To better understand how difficult each dimension was to annotate, we evaluated the "cognitive delay" related to the annotation of each dimension. We measured this delay by calculating the time interval between the affective event apex and the apex of the continuous time annotation curve for each dimension and participant, cf. Figure 5. After performing it on several selected extracts, average of difference time between annotators within each selected extract for each dimension was calculated. Extracts were selected on the basis of whether they described a bell-shaped curve, facilitating the apex identifications. The cognitive delay between dimension annotations was close, with the exception of the goal conduciveness annotation, for which the cognitive delay was longer. Results are reported in Table II.

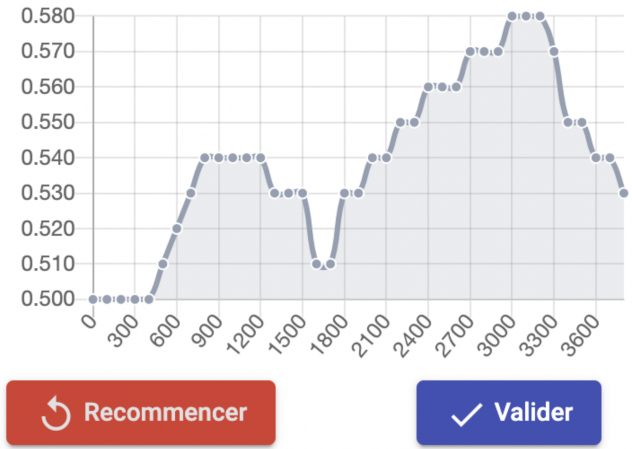

Fig. 2: Example of a time-continuous annotation of the arousal dimension; time is given in ms.

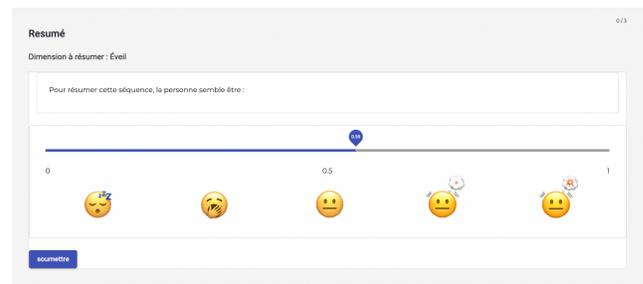

Fig. 3: Example of a summarised annotation of the arousal dimension.

### B. Relation between dimensions and affective label

The Figure 6 present a heatmap of the associations between affective labels and dimensions (summary) for each of the six annotators. Each cell of the heatmap corresponds to the value of the PCC calculated between a specific dimension and an affective label. Analysing the six heatmaps, we observe varying degrees of agreement in the correlations between appraisal dimensions and affective labels across the six annotators. For instance, no significant correlation was obtained between the label 'ashamed' and the five dimensions for annotators 3, 4 and 5. However, this label was correlated with at least one dimension for annotators 1, 2 and 3. On the opposite, the affective label 'surprise' shows consistently a strong correlations with the dimension of novelty for all six annotators, which indicates a reliable association. Conversely, the label 'sad' lacks significant correlation with any of the five dimensions for most annotators, showing its irrelevance in the context of AI-assisted CCT. Therefore, as explained in section V.C of the paper, affective labels that do not demonstrate significant correlations with appraisal dimensions for at least two annotators is considered one of the key selection criteria for excluding labels from the core set.

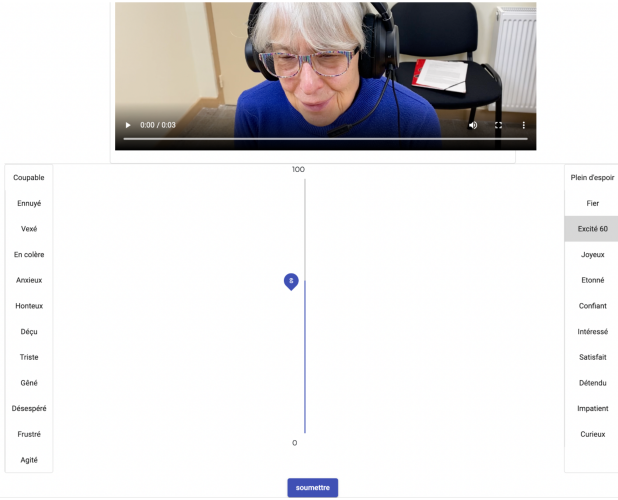

Fig. 4: Example of summary annotation of the labels.

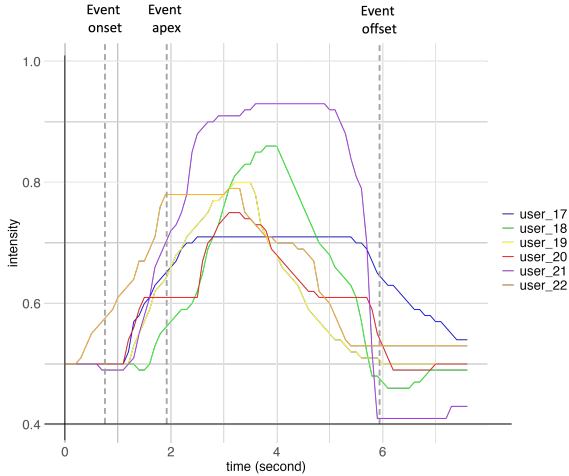

Fig. 5: An example of continuously valued annotation results on intrinsic pleasantness illustrating the cognitive delay of each annotator.

## IV. AUTOMATIC RECOGNITION OF AFFECTIVE STATES

### A. Representations of audio, textual and video data

In real-life applications such as THERADIA, human transcriptions are not always available. Therefore, the use of automatic transcription is required to experiment through the text modality. To this end, a benchmark was carried out to select the more robust ASR among those available. In accordance with an open science approach, the focus was placed on ASRs that are readily available and free of charge. The following candidates were selected: SpeechRecognition[2] toolkit, the Google, and the three Vosk[3] models that were trained on French speech (using the Kaldi toolkit[4]).

To compare the ASR models, Word Error Rate (WER) was calculated on all of the annotated sequences (see Table III). The Google ASR achieved the best results (33.69 %

TABLE II: Cognitive delay in continuous annotation as a function of dimensions

|  | Arousal | Novelty | Intrinsic pleasantness | Goal conduciveness | Coping |
|---|---|---|---|---|---|
| **Number of sequences** | 4 | 6 | 5 | 5 | 7 |
| **Cognitive delay (s)** | 1.99 ± 0.43 | 1.32 ± 0.51 | 1.13 ± 0.36 | 3.30 ± 0.65 | 1.33 ± 0.48 |

average WER), followed by the Vosk-linto model (38.76 % average WER). Thus, automatic transcriptions of the current experiment were performed using the Google ASR.

The performance of predicting the intensity of affective labels of the core set was compared based on human or automatic transcriptions, using TF-IDF and BERT features. The results are reported in table IV. They suggest a small drop in performance when using automatic transcriptions instead of human transcription. Specifically, similar performance was revealed between human and automated transcriptions with the use of TF-IDF ($p > .05$ based on two tailored z-test). However, a small effect of better performance was evidenced for human compared to automated transcription with the use of BERT (Cohen's $d = 0.25$).

Given the small drop in performance compared with human transcription, automatic transcription seems suitable for predicting affective labels. It should also be noted that these results are in line with the literature [1]–[4].

### B. Prediction bias based on population demographics

A prediction bias analysis of affective labels and dimensions was performed based on population demographics, i.e., male vs. female; healthy seniors vs. MCI. To this end, the test partition of the corpus was divided according to the two groups of participants, i.e., seniors and MCI, then the seniors were divided according to the gender. This analysis was performed on multimodal predictions of time-continuous dimensions using the EWE as gold standard, the average of dimension summaries, and the average of label intensities.

The results for the prediction of label intensity and dimensions are reported in table V, and table VI, respectively. Based on two-tailored z-test, both expert and deep representations show better performance for the female group than for the male group, $p < .05$. This could be explained by the female to male ratio in the participant sample (40/12), which translates into more robust training for the female group. Moreover, significantly better performances were evidenced for the MCI group compared to the healthy senior group, $p < .05$. This result was unexpected, given that the ratio of MCI to healthy seniors is low in the sample of participants (9/41), leaving less data for the models to train on. It is possible that MCI participants expressed more pronounced affects than healthy seniors facilitating the model training and prediction. The rationale of this assumption is underpinned by the fact that they were likely more involved in the CCT since they were directly concerned by the cognitive remediation aspect.

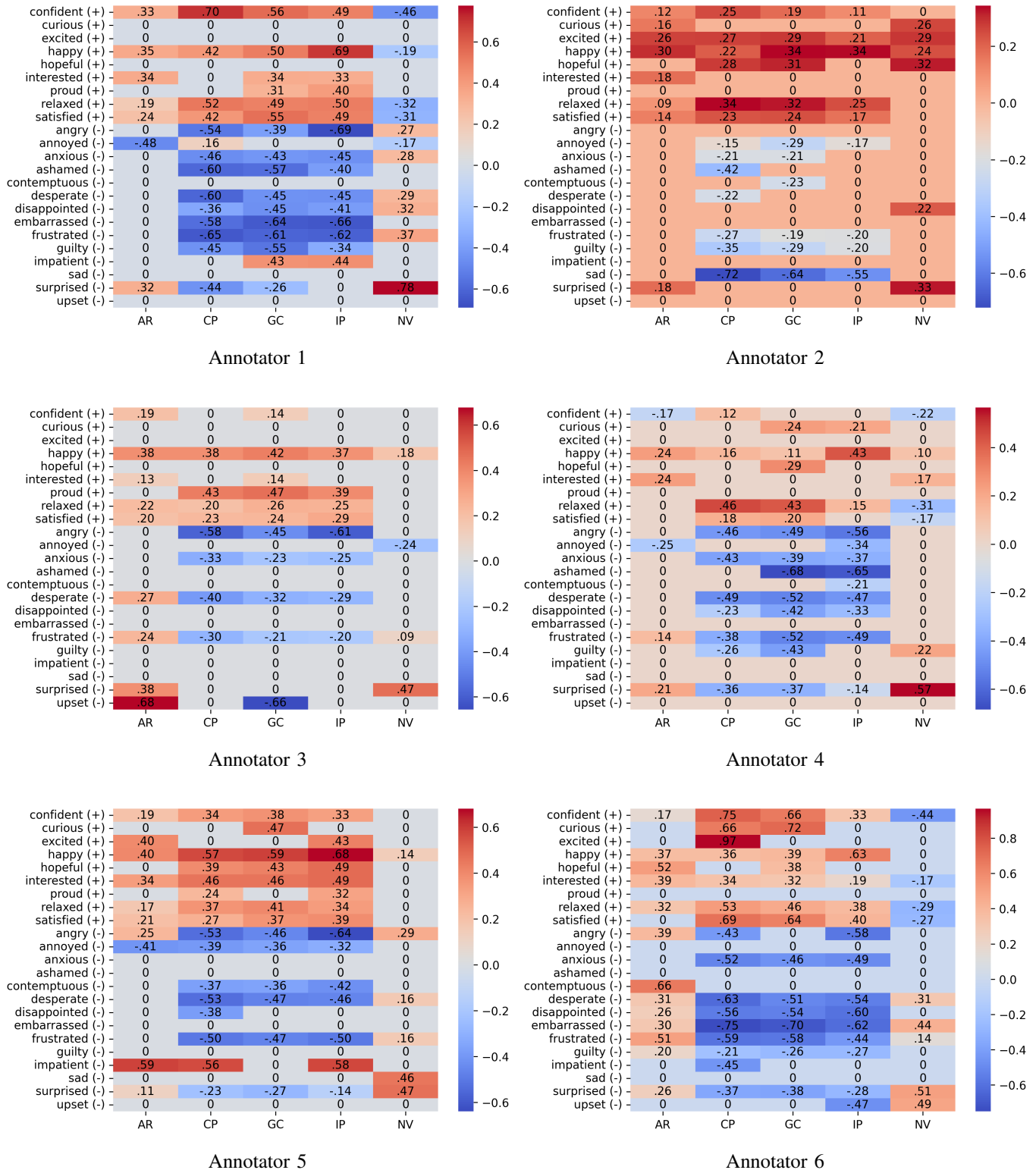

Fig. 6: Heatmaps of the PCC between each dimension (summary value) presented in the columns – AR: arousal, CP: coping, GC: goal conduciveness, IP: intrinsic pleasantness, and NV: novelty – and each affective label on the rows, which are sorted in alphabetical order from positive (+) to negative (-), for the evaluations from each of the six annotators. Significant correlations (p-value $\leq 0.05$) are coloured either in red for positive values, or blue for negative values. Non-significant correlations (p-value $> 0.05$) are presented with 0 shown in grey.

TABLE III: Automatic Speech Recognition (ASR) results on all the transcribed sequences, where there exists text based on human transcription. The results are mean (std) of Word Error Rate (WER) for all sequences according to different ASR models

| ASR model | Senior | | | MCI | | | Senior + MCI | | |
|---|---|---|---|---|---|---|---|---|---|
| | Emotional | Neutral | All | Emotional | Neutral | All | Emotional | Neutral | All |
| Google | 55.45% (43.40%) | 49.48% (44.96%) | 54.15% (43.82%) | 49.56% (39.83%) | 46.72% (43.98%) | 47.80% (42.45%) | 55.22% (43.28%) | 48.95% (44.78%) | 53.65% (43.74%) |
| Vosk-small | 57.79% (43.97%) | 52.60% (44.84%) | 56.65% (44.21%) | 48.67% (39.45%) | 40.33% (41.51%) | 43.50% (40.92%) | 57.43% (43.83%) | 50.23% (44.47%) | 55.63% (44.10%) |
| Vosk-medium | 48.00% (49.05%) | 38.49% (45.49%) | 45.92% (48.45%) | 39.82% (42.72%) | 29.04% (39.60%) | 33.14% (41.12%) | 47.68% (48.84%) | 36.67% (44.57%) | 44.92% (48.04%) |
| Vosk-linto | 56.51% (44.66%) | 50.90% (44.94%) | 55.28% (44.78%) | 49.10% (39.42%) | 40.04% (41.12%) | 43.48% (40.70%) | 56.21% (44.49%) | 48.81% (44.43%) | 54.36% (44.58%) |
| Frequency | 7859 | 2201 | 10060 | 322 | 526 | 848 | 8181 | 2727 | 10908 |
| Word / second | .791 (.624) | .719 (.563) | .775 (.611) | .877 (.641) | .748 (.601) | .797 (.620) | .794 (.625) | .725 (.570) | .777 (.612) |

TABLE IV: Affect label intensity prediction from text using either human or automatic transcriptions (CCC).

| Label | TF-IDF | | BERT | |
|---|---|---|---|---|
| | Human | Automatic | Human | Automatic |
| Annoyed | .031 | .075 | .122 | .255 |
| Anxious | .140 | .115 | .124 | .087 |
| Confident | .263 | .233 | .518 | .473 |
| Desperate | .211 | .172 | .322 | .221 |
| Frustrated | .381 | .327 | .183 | .222 |
| Happy | .127 | .117 | .157 | .108 |
| Interested | .156 | .147 | .336 | .364 |
| Relaxed | .341 | .314 | .295 | .363 |
| Satisfied | .344 | .323 | .364 | .057 |
| surprised | .209 | .167 | .197 | .143 |
| Average | .220 | .199 | .262 | .229 |

TABLE V: The bias and fairness analysis of the multi-modal emotion regression models. The reported results are in Concordance Correlation Coefficients (CCC).

| Label | Traditional features | | | | | Deep representations | | | | |
|---|---|---|---|---|---|---|---|---|---|---|
| | Female | Senior Male | F+M | MCI | Senior+MCI | Female | Senior Male | F+M | MCI | Senior+MCI |
| Annoyed | .210 | .131 | .247 | .367 | .255 | .302 | .199 | .284 | .328 | .280 |
| Anxious | .202 | .099 | .174 | .128 | .171 | .234 | .087 | .209 | .266 | .214 |
| Confident | .434 | .361 | .424 | .337 | .422 | .498 | .358 | .477 | .575 | .487 |
| Desperate | .285 | .178 | .260 | .218 | .259 | .249 | .252 | .248 | .333 | .252 |
| Frustrated | .472 | .143 | .377 | .508 | .387 | .478 | .403 | .463 | .491 | .465 |
| Happy | .456 | .402 | .471 | .597 | .490 | .508 | .455 | .508 | .523 | .508 |
| Interested | .208 | .179 | .210 | .212 | .207 | .269 | .336 | .295 | .374 | .300 |
| Relaxed | .492 | .529 | .510 | .548 | .515 | .520 | .475 | .516 | .646 | .529 |
| Satisfied | .439 | .351 | .435 | .365 | .428 | .490 | .496 | .499 | .348 | .486 |
| Surprised | .204 | .072 | .179 | .339 | .197 | .284 | .146 | .268 | .327 | .275 |
| Average | .340 | .244 | .329 | .362 | .333 | .383 | .321 | .377 | .421 | .380 |

TABLE VI: The bias and fairness analysis of the multi-modal dimension regression models. The reported results are in Concordance Correlation Coefficients (CCC).

| Dimensions | Time-continous | | | | | Summaries | | | | |
|---|---|---|---|---|---|---|---|---|---|---|
| | Female | Senior Male | F+M | MCI | Senior+MCI | Female | Senior Male | F+M | MCI | Senior+MCI |
| Arousal | .695 | .638 | .682 | .777 | .691 | .280 | .361 | .315 | .360 | .310 |
| Coping | .430 | .328 | .406 | .511 | .418 | .690 | .608 | .674 | .799 | .684 |
| Goal conduciveness | .422 | .382 | .413 | .490 | .420 | .668 | .610 | .658 | .711 | .662 |
| Intrinsic pleasantness | .392 | .281 | .366 | .404 | .370 | .634 | .448 | .598 | .709 | .609 |
| Novelty | .351 | .242 | .325 | .455 | .339 | .607 | .348 | .543 | .709 | .559 |



\IEEEpubid{0000--0000/00\$00.00~\copyright~2021 IEEE}


\vfill

\end{document}